\documentclass[trackchanges, twocolumn]{aastex7}

\usepackage{amsmath}
\usepackage{enumitem}
\usepackage{float}
\usepackage{hyperref}

\begin{document}

\title{Galactic bars are already mature at Cosmic Noon: bar strength and flatness at $z \sim 1.5$}

\correspondingauthor{Boris S. Kalita}
\email{boris.kalita@pku.edu.cn; kalita.boris.sindhu@gmail.com}

\author[0000-0001-9215-7053]{Boris S. Kalita}
\altaffiliation{Kavli Astrophysics Fellow and Boya Fellow}
\affiliation{Kavli Institute for Astronomy and Astrophysics, Peking University, Beijing 100871, People's Republic of China}
\affiliation{Kavli IPMU (WPI), UTIAS, The University of Tokyo, Kashiwa, Chiba 277-8583, Japan}
\email{boris.kalita@pku.edu.cn; kalita.boris.sindhu@gmail.com}

\author[0000-0001-6947-5846]{Luis C. Ho}
\affiliation{Kavli Institute for Astronomy and Astrophysics, Peking University, Beijing 100871, People's Republic of China}
\affiliation{Department of Astronomy, School of Physics, Peking University, Beijing 100871, People's Republic of China}
\email{lho@pku.edu.cn}

\author[0000-0002-0000-6977]{John D. Silverman}
\affiliation{Kavli IPMU (WPI), UTIAS, The University of Tokyo, Kashiwa, Chiba 277-8583, Japan}
\affiliation{Department of Astronomy, School of Science, The University of Tokyo, 7-3-1 Hongo, Bunkyo, Tokyo 113-0033, Japan}
\affiliation{Center for Astrophysical Sciences, Department of Physics \& Astronomy, Johns Hopkins University, Baltimore, MD 21218, USA}
\email{john.silverman@ipmu.jp}

\author[0000-0002-5743-0250]{Frédéric Bournaud}
\affiliation{CEA, Irfu, DAp, AIM, Universit\`e Paris-Saclay, Universit\`e de Paris, CNRS, F-91191 Gif-sur-Yvette, France}
\email{frederic.bournaud@cea.fr}

\author[0000-0003-0348-2917]{Miroslava Dessauges-Zavadsky}
\affiliation{Department of Astronomy, University of Geneva, 51 Chemin Pegasi, 1290 Versoix, Switzerland}
\email{miroslava.Dessauges@unige.ch}

\author[0000-0002-3331-9590]{Emanuele Daddi}
\affiliation{CEA, Irfu, DAp, AIM, Universit\`e Paris-Saclay, Universit\`e de Paris, CNRS, F-91191 Gif-sur-Yvette, France}
\email{emanuele.daddi@cea.fr}

\author[0000-0001-9369-1805]{Annagrazia Puglisi}
\altaffiliation{Anniversary Fellow}
\affiliation{School of Physics and Astronomy, University of Southampton, Highfield SO17 1BJ, UK}
\email{a.puglisi@soton.ac.uk}

\author[0000-0002-0786-7307]{Xuheng Ding}
\affiliation{School of Physics and Technology, Wuhan University, Wuhan 430072,  China}
\email{dingxh@whu.edu.cn}

\author[0000-0002-3462-4175]{Si-Yue Yu}
\affiliation{Kavli IPMU (WPI), UTIAS, The University of Tokyo, Kashiwa, Chiba 277-8583, Japan}
\email{si-yue.yu@ipmu.jp}




\begin{abstract}

In this work, we explore the nature of $z>1$ galactic bars. Once thought to be highly transient, our results demonstrate otherwise. Our sample consists of nine massive ($>10^{10.5}\,\rm M_{\odot}$) star-forming barred-spiral galaxies at $z_{\rm spec} \sim 1.5$. Using rest-frame near-IR (F444W) JWST/NIRCam imaging, we apply ellipse fitting along with 1D and 2D morphological modeling to directly measure bar properties. We find that five galaxies host flat surface brightness profiles (bar Sérsic index $<0.4$), indicative of highly evolved, ``mature" bars.  By contrast, only two galaxies show exponential profiles, characteristic of young bars, and these are also shorter in absolute length than the flat bars.  We therefore conclude that a large fraction of bars at this epoch have already matured, thereby indicating the presence of well-settled disks required to facilitate bar formation and sustained evolution well before $z\sim1.5$. To assess the gravitational impact of the bars, we calculate the maximum transverse-to-radial force ratio ($Q_{b}$). We find that $Q_{b}$ values are comparable to, or weaker than, those of bars in the local Universe,  Seven of the nine bars show only a marginal increase in strength with maturity (from exponential to flat bars). Contrarily however, the remaining two bars are flat,  but have the lowest $Q_{b}$ values in our sample.  We hence propose that the mature bars at $z\sim 1.5$ may experience phases of weakening due to rapid gas inflows and/or minor mergers. In conclusion, our work sheds light on the rapidly evolving nature of high-z bars and paves the way for larger statistical studies.

\end{abstract}



\section{Introduction} \label{sec:intro}

The galactic bar is a ubiquitous morphological feature, with the fraction of galaxies hosting a bar steadily increasing from $\sim 10-20\%$ at $z \sim 1$ to $\sim 70\%$ in the local Universe \citep[e.g.,][]{sheth03, jogee04, elmegreen04, marinova07, sheth08, masters11, simmons14, rosas-guevara20,zhao20,leconte24, guo25, euclid25, geron25,salcedo25,pastras25}. These elongated structures, consisting of stable stellar orbits spanning the inner regions of disk galaxies, arise from growing non-axisymmetric instabilities in dynamically cold disks, with a Toomre-$Q$ parameter $\approx 1.5-2.0$ \citep{combes81, combes93, athanassoula13b}. Once formed, these instabilities grow as stars lose angular momentum and join the elongated orbits making up the bar \citep{semczuk24}. Theoretical works suggest that dynamical friction with the dark matter halo \citep{athanassoula02, athanassoula13} and high-angular-momentum gas accretion to the disk outskirts \citep{bournaud05c} regulate bar strength. Tidal interactions have also been proposed to trigger bar formation and evolution \citep{gerin90, noguchi96, miwa98, berentzen04, romanodiaz08, lokas14, martinezvalpuesta17, lokas18, peschken19, ansar25, zheng25}. Attempts to verify this interaction-driven pathway observationally remain inconclusive in the local Universe \citep{vandenbergh02, aguerri09, barway11, lee12}. However, tidally induced bars may dominate at $z \gtrsim 2$ \citep{bi22, tsukui24}, although \cite{huang25} argue the opposite for a barred galaxy at $z \sim 2.5$.

The growth of bars has also been characterized through changes in the surface brightness profiles along the major axis of the bar \citep{gadotti11, anderson22, erwin23}.  The pioneering study of \cite{elmegreen85} divided bars into two profile types: ``flat" and ``exponential".  Parameterizing these profiles with Sérsic models, \cite{kim15} further subdivided flat bars into flat ($n < 0.4$) and intermediate ($0.4 < n < 0.8$), using the Spitzer Survey of Stellar Structure in Galaxies \citep[S$^{4}$G;][]{sheth10}. Meanwhile, $n > 0.8$ profiles are defined as exponential. Bars are expected to initially follow the exponential disk profile, and are therefore more common in low-mass, star-forming ``late-type" systems \citep{combes93, kruk18}. In contrast, massive and evolved ``early-type" systems host flat, mature bars \citep[also called``peak" and ``shoulder" profiles, with the shoulders resulting in the characteristic flatness;][]{erwin23}. $N$-body simulations show that as bars grow, particles are trapped in looped $x_{1}$ orbits, producing flat profiles with shoulders. These shoulders may first appear as temporary structures at $\sim 1-2\,\rm Gyr$ after bar formation, but become stable after $\gtrsim 2-3\,\rm Gyr$ \citep{anderson22, beraldo23}. However, \cite{noguchi96} suggested that shoulders, and hence flatness, can also result from tidal interactions. This aligns with studies indicating that tidally induced bars may prematurely evolve into mature states \citep{miwa98, fragkoudi25, zheng25}.

The timing of local bar formation is linked to the epoch of ``disk settling", originally expected at $z \gtrsim 1$ \citep{sheth08, kraljic12, kim15, kim21}. However, the exact epoch remains unclear, with observations finding indications of bar-formation as early as $z \sim 3.5-4.0$ \citep{smail23,constantin23,amvrosiadis25,geron25,guo25,leconte25}.  These may be lower limits as resolution effects would likely limit detection of early phases of bar formation. With bars now detected at such high redshifts, their nature becomes the key question: are they long-lived or transient structures? Do they exert a dynamical effect on the host disk comparable to that in the local Universe? 

This work explores an alternate way of answering these questions, by measuring the major-axis surface brightness profiles of bars at $z \sim 1.5$. We aim to determine if they are flat and ``mature", having persisted for $\gtrsim 1-2\,\rm Gyr$ \citep[indicated by N-body simulations, without the inclusion of gas and star-formation;][]{anderson22},  and therefore since $z\sim2.5-3.5$.  Alternatively, if the bars exhibit exponential profiles,  it would imply repeated destruction and reformation due to tidal disruptions and/or dissipation from high gas densities \citep[e.g.,][]{elmegreen04, elmegreen05b, kraljic12}. We also quantify their strengths and impact on their hosts by calculating the transverse force exerted by the bar potential, enabling a direct comparison of bar strengths at $z \sim 1.5$ with those in the local Universe \citep{lee20}.

\section{Sample} \label{sec:sample}
We investigate the profiles of bars in nine barred-spiral galaxies at $z \sim 1.5$. These galaxies, identified as barred (selection described in Sec.~\ref{sec:analysis}), are part of the the sample presented in \cite{kalita25a} (BK25 hereafter). This parent sample of 32 star-forming main-sequence galaxies were selected to be face-on (disk axis ratio $>0.6$) and without clear major-merger signatures in rest-frame near-IR imaging. Each galaxy is spectroscopically confirmed to be within $1.4 \leq z \leq 1.7$ \citep[as part of the FMOS H$\alpha$-detected sample in the COSMOS field;][]{silverman15,kashino19}, and has a stellar mass within the range of $10^{10.5-11.4}\,\rm M_\mathrm{\odot}$. The sample also has multi-band (F115W, F150W, F277W, and F444W) JWST/NIRCam coverage from the COSMOS-Web survey \citep{casey23}. 

The goal of this study is the detection of bars, which are most consistently identified at $z > 1$ in rest-frame near-IR wavelengths \citep{guo23}. For our sample, BK25 likewise found that stellar substructures such as spirals and bars are detected mainly in the F444W image ($\sim 1.77\,\mu m$ rest-frame at $z \sim 1.5$). We therefore restrict our analysis to bars in the F444W images, while a joint multi-wavelength analysis will be presented in future work.

\section{Morphological Analysis} \label{sec:analysis}
In the following subsections, we discuss the set of analyses we undertake on the F444W images to derive the structural properties of the galactic bars in our sample.
\subsection{Ellipse fitting} \label{subsec:ellipse_fit}
\begin{figure*}
    \centering
    \includegraphics[width=\textwidth]{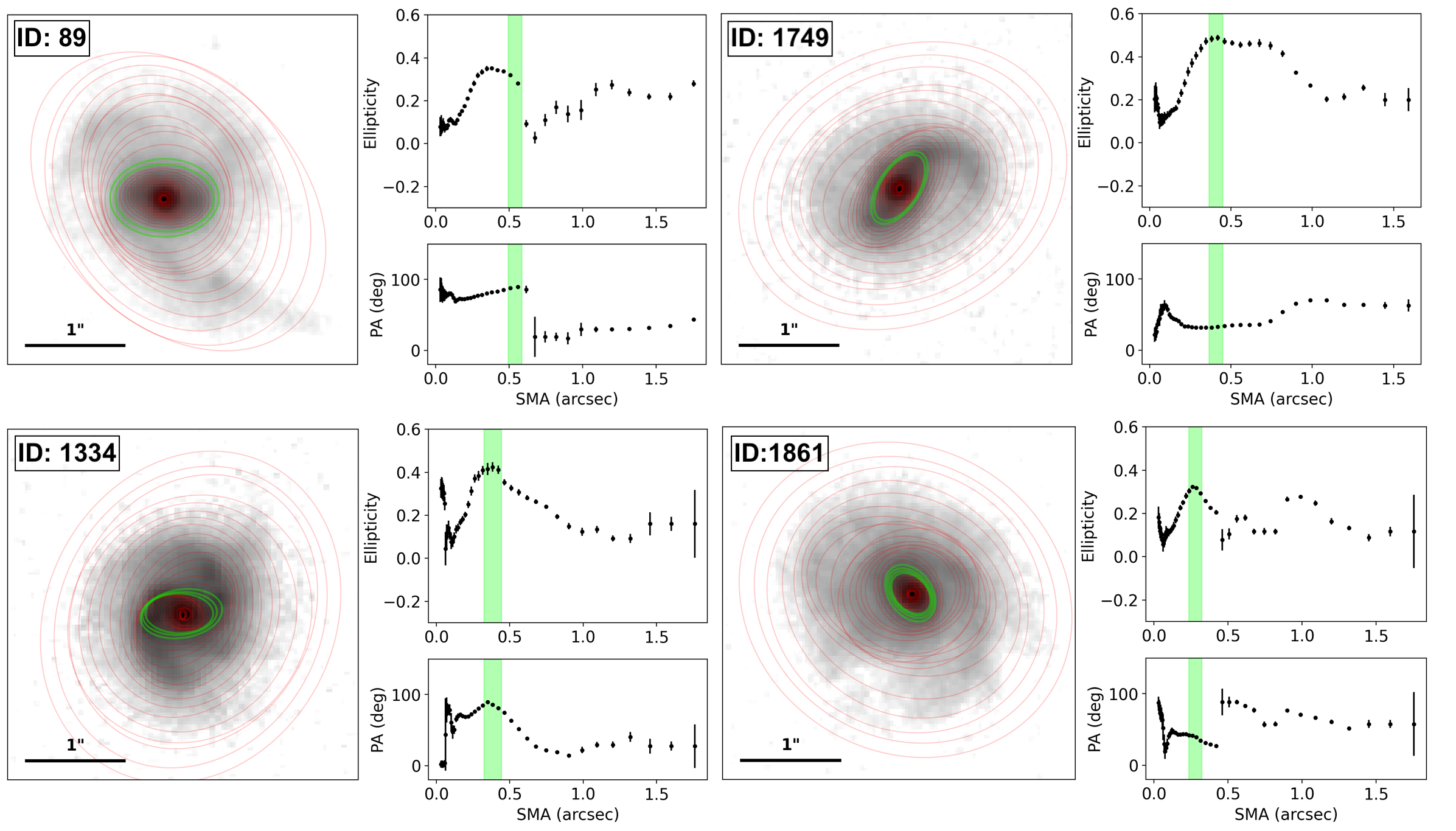}
    \caption{Ellipse fitting: (for each galaxy; left) The F444W image with ellipses fit to constant surface brightness isophotes.  The ellipse with SMA $=$ bar length is highlighted.  (top right) The ellipticity vs SMA plot. The SMA value at the peak ellipticity is used as the bar length. The exception being ID: 89 (and ID: 1147, not shown), which rather shows a plateau. In this case we select the outer edge of this plateau as the peak since it also corresponds to a change in the position angle (PA) change of $>10^{\circ}$ right after the fall of the peak in ellipticity. This can be seen in the PA vs SMA plot (bottom right).}
    \label{fig:ellipse_fit}
\end{figure*}
When observing the galactic disk face-on, stellar bars are non-axisymmetric structures with centers consistent with the host disk\footnote{However, off-centered bars have also been observed \citep[][]{kruk17,deswardt15}}. Therefore, the higher ellipticity ($\equiv 1-$axis ratio) of the bar compared to the disk can be used to systematically identify the presence of a bar. The commonly used method involves fitting radially concentric ellipses to the flux isophotes of the image, with the expectation that the ellipses with the highest ellipticity correspond to the bar \citep[e.g.,][]{wozniak95, jogee04, marinova07, yu22,salcedo25}. This procedure has now also been implemented at $z>1$, although visual inspection is still necessary to ensure proper bar identification \citep{guo23,guo25,leconte24,leconte25}.

We implement this method on the BK25 sample, with two examples shown in Fig.~\ref{fig:ellipse_fit}.  We begin with the F444W image. Using the \textsc{photutils}\footnote{\textsc{https://photutils.readthedocs.io/}}
 \citep{photutils} package, we fit progressively larger ellipses to fainter isophotes (Fig.~\ref{fig:ellipse_fit}, left in each panel). We start at the center determined from the bulge-disk F444W image morphological modeling in BK25,  and a semi-major axis (SMA) of 2 pixels ($\sim$ F444W PSF FWHM in pixels).  After the first ellipse is fit to the isophote at SMA $=$ 2 pixels,  the code incrementally increases the SMA by a factor of $0.1$ and fits the next isophote. We allow the progressive ellipse centers to shift freely to enable the detection of offset bars. The ellipticity (Fig.~\ref{fig:ellipse_fit},top right, in each panel) and position angle (PA; Fig.~\ref{fig:ellipse_fit}, bottom right, in each panel) of the series of ellipses can thus be plotted as a function of their SMA.  A highly elliptical structure like a bar will appear as a local maxima in the ellipticity vs.  SMA plot,  beyond which the larger ellipses fit isophotes representing the more circular disk structure. 

We then locate the peak in the ellipticity vs. SMA plot \citep[local maxima with an increase $>0.1$ in ellipticity; condition defined in][]{marinova07} within the disk $r_\mathrm{90}$ (radius of $90\%$ flux), also from the bulge-disk modeling in BK25.  If the peak occurs at SMA values $<$ PSF FWHM/2, we reject it, as \cite{liang24} found bar detectability decreases drastically below this threshold. Given our small sample of 32 objects, we use visual inspection to ensure that ellipses corresponding to the peak ellipticity represent bars and not other disk substructures, such as spiral arms and clumps, which can be clearly identified. We find these substructures produce secondary spikes in ellipticity (e.g., ID 1749 and 1861,  both shown in Fig.~\ref{fig:ellipse_fit}), as has also been noted in other $z>1$ studies \citep[e.g.,][]{guo23}.  

A second criteria is that of the PA being constant \citep[$<10^{\circ}$;][]{kim21} across the length of the bar, since it is a structure with a fixed rotational velocity. However,  to do this, we have to determine the inner radius of the bar (or at least a range of SMA datapoints clearly tracing the bar) to define a region over which the PA would remain constant. The current resolution and depth prevents us from satisfactorily doing that for most of the sample.  However, in cases of (visually) strong bars, the peak ellipticity appears as a plateau with a constant PA throughout. In these cases, we invoke the PA criteria and define the bar length as the SMA value within this plateau that also corresponds to a decrease in the phase angle $>10^{\circ}$ following the drop in ellipticity within $~\sim 0.03^{\prime\prime}$  (the pixel size; see ID 89 in Fig.~\ref{fig:ellipse_fit}). 

In all cases where no bar is found using our procedure, we exclude the galaxy from our sample. Applying this to all 32 galaxies in BK25, we identify nine with bars. This constitutes our final sample. The SMA value corresponding to the peak is considered the bar length.

\subsection{Masking the spiral arms} \label{subsec:spiral_mask}
The primary goal of this work is to model the profiles of the bars (Sec.~\ref{subsec:1d_fit} and Sec.~\ref{subsec:2d_fit}). We find that the presence of spiral arms interferes with such analyses, as indicated by secondary spikes in the ellipse-fitting method (Sec.~\ref{subsec:ellipse_fit}). 
The presence of spiral arms in the sample galaxies have already been commented on in BK25 and a subset of these have been analysed in \cite{kalita25c}.  In the previous section we find that these structures interfere with the ellipse-fitting method by increasing the ellipticity of isophotes.  Therefore, in this section, we create masks for the spiral arms in preparation of the image-based morphological analysis of the bars (Sec.~\ref{subsec:1d_fit} and Sec.~\ref{subsec:2d_fit}). To achieve this, we create 2D masks of the logarithmic spirals in these galaxies using a polar-coordinate wavelet decomposition method developed in \cite{kalita25c}. This method involves de-projecting the galaxy using the disk axis ratio (from the BK25 bulge-disk modeling), followed by a wavelet decomposition in polar ($r,\theta$) coordinates. Using the $m=2-6$ modes (signals that repeat $2-6$ times over $\theta = 0-2\pi$), a ``spiral-like" image is constructed that contains the spiral structures, if present in a galaxy. 

In \cite{kalita25c}, this step was followed by fitting a spiral galaxy model using \textsc{GALFIT} \citep{peng02,peng10}. However, this was only possible for six out of 32 galaxies, mainly due to additional disk substructures complicating the fit. We have now developed a simpler tool to ``detect" logarithmic spirals in the spiral-like image by fitting simple spiral models.  The parametric function in polar co-ordinates for each logarithmic spiral arm can be defined as:

\begin{equation*}
r(\theta) = a \, e^{b\theta},
\end{equation*}
where \(a\) is a scaling constant controlling the tightness of the spiral, and $b = \tan(\phi)$,with $(\phi)$ being the pitch angle of the spiral in radians. Converting to Cartesian co-ordinates, we get:
\begin{equation} \label{eq:skeleton_eq}
\begin{split}
x(\theta) = r(\theta) \cos(\theta + \Delta) + x_c, \\
y(\theta) = D.\, r(\theta) \sin(\theta + \Delta) + y_c
\end{split}
\end{equation}
Here the additional factor of $\Delta$ (in radians) is introduced, which controls the starting phase angle of each spiral and is helpful while fitting multiple arms. The parameters $x_c, y_c$ are the galactic center. Finally the `$D$' parameter which can be values either $-1$ or $1$ controls whether a spiral arm propagates clockwise or anti-clockwise.  With each arm path defined using Eq.~\ref{eq:skeleton_eq}, and with a thickness of 5 pixels ($\sim$ PSF size), we can generate multi-arm spiral `skeletons' over a parameter space defined by the number of arms, $\Delta$ and $\phi$. These skeletons are fit to the aforementioned spiral-like image and we optimise the parameters by maximising the contrast between the flux within the skeletons and the rest of the galaxy. Once the final spiral skeleton is identified, we create a mask by thickening each arm by 9 pixels \citep[$\sim 2\,\rm kpc$ at $z = 1.5$, the approximate width of the arms in the sample;][]{kalita25c}. We then undo the de-projection to obtain the mask for the original galaxy image.


Finally, we remove any part of the spiral arm mask that falls within the ellipse corresponding to the bar (Sec.~\ref{subsec:ellipse_fit}).  This may happen since the bar will be included in the spiral-like image we use to generate the masks, since it is also a structure dominant in the $m=2$ mode. However, we modify this step for two of the galaxies (IDs: 89 and 1147) which feature clearly visible strong bars.  As discussed in the previous section, the ellipticity vs. SMA plot of such galaxies show a plateau with an inner peak as well as an outer peak. Although we use the outer peak to fix the bar ellipse, we rather use the inner peak to determine the minimum radius where we mask the spiral arms. We take this conservative approach since the flagging occurs only on one side of the bar in both cases, and hence there is still sufficient flux for the proper surface brightness modelling. 

\subsection{1-D major axis profile} \label{subsec:1d_fit}
The profile of a galactic bar can be assessed using the flux distribution across the galaxy along the bar SMA \citep[bar-major-axis profiles, mainly applied for galaxies in the local Universe; e.g.,][]{erwin23, gusev03}. This is commonly used to determine whether the bar follows the exponential radial light profile of the disk\footnote{Disks are expected to be exponential even at $z>1$ \citep{elmegreen05b}} (exponential bars) or exhibits the characteristic shoulders leading to a flatter appearance (flat–intermediate bars). Once a bar is detected in a galaxy in our sample, we generate similar bar-major-axis profiles by using the position angle of the ellipse corresponding to each bar (Sec.~\ref{subsec:ellipse_fit}) to define a long slit-like aperture of thickness $=5$ pixels ($\sim$ PSF FWHM) along the bar.
\begin{figure}
\centering
\includegraphics[width=0.49\textwidth]{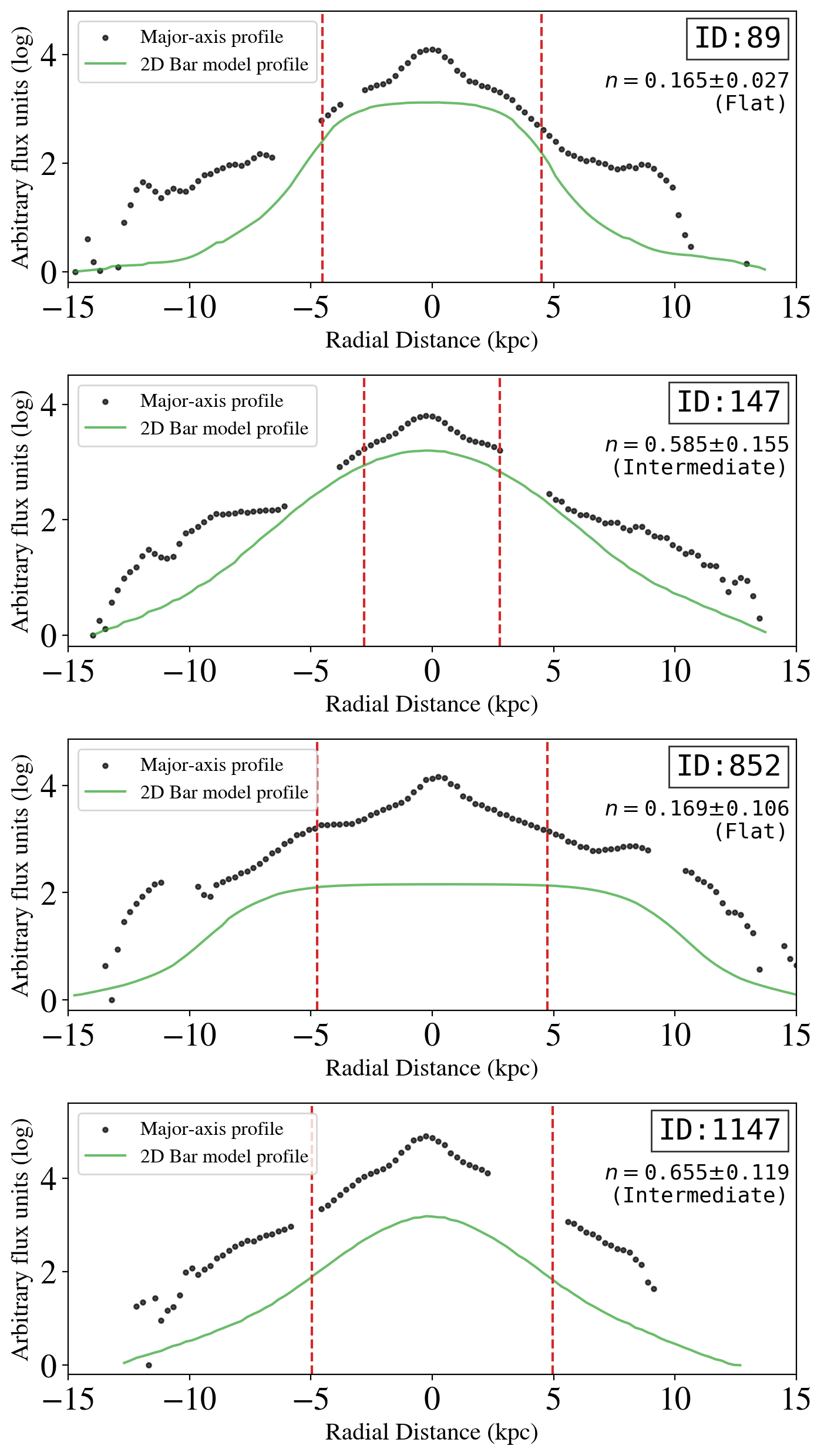}
\caption{The 1D surface brightness profiles along the bar major axis of the galaxies in our sample (continued in Fig.~\ref{fig:profile_examples_2}). Also shown are the major-axis profiles of the 2D best fit bar models\footnote{the 1D profiles of the 2D bar models are generated through the same procedure that was used to generate the 1D profiles from the galaxy images in Sec.~\ref{subsec:1d_fit}} (Sec.~\ref{subsec:2d_fit}) for comparison. The x-axis gives the radial distance in kpc from the galaxy center. The red dashed lines denote the $R_\mathrm{bar}$, which is considered the approximate extent of the bar.  Hence, the region within these lines contain the majority of the bulge and bar flux.  In each case, we also provide the bar $n$-value from the 2D morphological fitting (Sec.~\ref{subsec:2d_fit}), as well as the resulting classification into flat, intermediate and exponential bar profiles.  In the five galaxies with flat profiles, the characteristic shoulders beyond the central bulge can be seen, giving the appearance of a ``flattening" of the profile within $R_\mathrm{bar}$. The gaps in the profile are regions which have been flagged as they likely are regions influenced by spiral arms (Sec.~\ref{subsec:spiral_mask}).}
\label{fig:profile_examples}
\end{figure}

\begin{figure}
\centering
\includegraphics[width=0.49\textwidth]{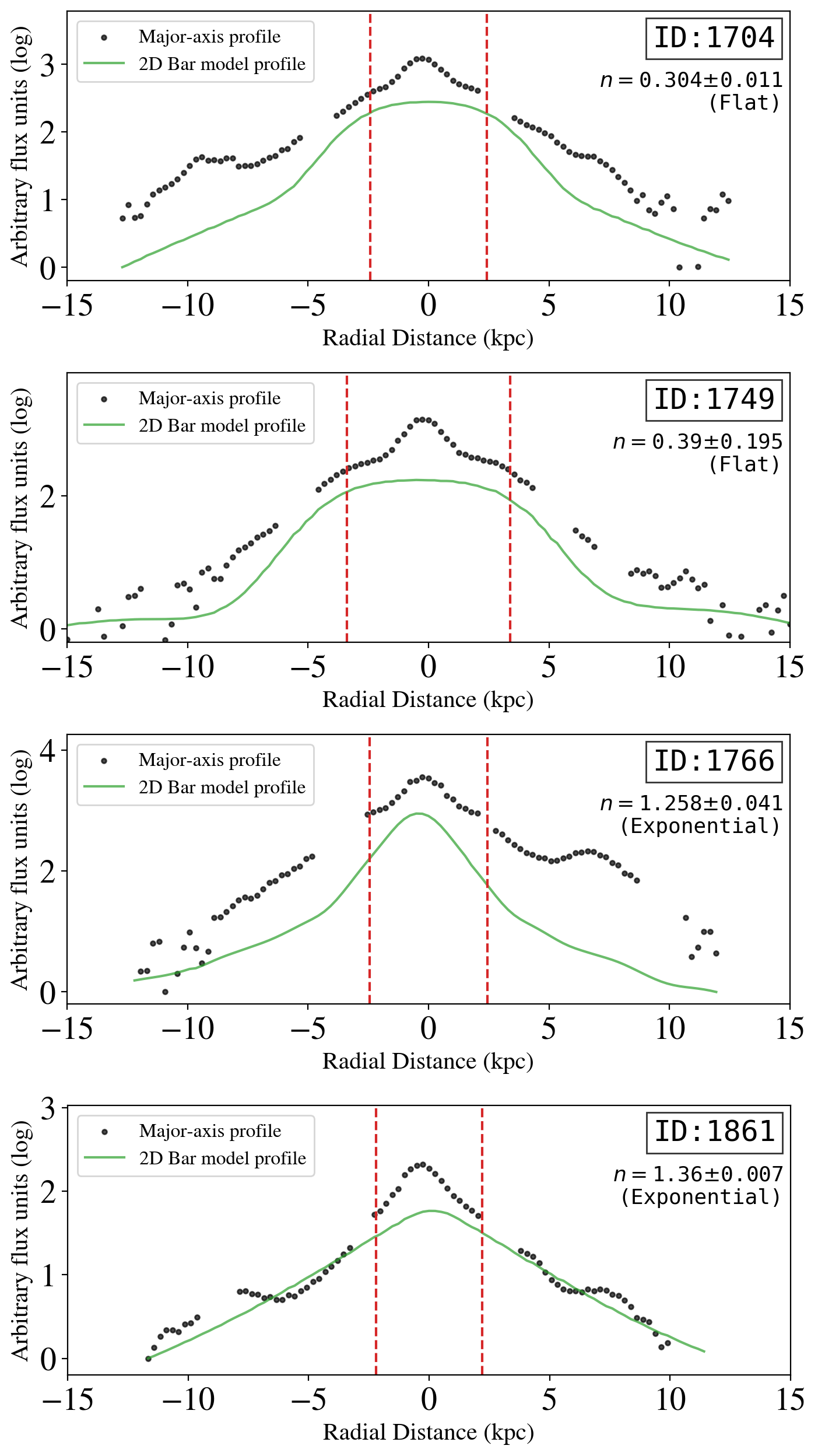}
\caption{Continuation of Fig.~\ref{fig:profile_examples} for the remaining galaxies.}
\label{fig:profile_examples_2}
\end{figure}

Collapsing the slit aperture over its width yields the 1-D flux distribution, which contains contributions from the bulge, bar, and disk, with spiral arms already masked out. The resulting major-axis profiles, shown in Fig.~\ref{fig:profile_examples}, are then fit with a composite model consisting of three Sérsic (F444W PSF-convolved) components: bulge ($n=4$), disk ($n=1$), and bar ($n=0.1-1.5$).  Details of the PSF generation is mentioned in the next section. We decide on the Sérsic index range for the bar based on the local Universe study by \cite{kim15}. We find it to be the largest study that implements fitting of Sérsic profiles to the surface brightness of bars.  Hence we implement the same range in this study, not just for the 1D profile fitting, but also for the 2D fitting discussed next.  The fit is first optimized using the Python package \textsc{SCIFIT}, followed by an MCMC $\chi^2$-optimization procedure. Although this 1-D morphological fitting is less robust than a 2-D approach—since it cannot fully exploit structural dissimilarities between the components—it still serves as a method of characterizing the distinct bar major-axis profiles observed visually.

\subsection{2-D morphological fitting}\label{subsec:2d_fit}
\begin{figure*} 
    \centering
    \includegraphics[width=0.96\textwidth]{2D_comp_1.png}
    \caption{The 2D morphological fitting of the rest-frame near-IR F444W images for six of the nine galaxies in our sample. The bulge, disk and bar models are shown separately, along with the residual $ = $ Image $-$ (bulge+disk+bar) model. The final image is scaled based on the signal-to-noise (S/N), which is the ratio of the residual flux and the noise image.}
    \label{fig:2D_fit_comp_1}
\end{figure*}
\begin{figure*}[h]
\centering
\includegraphics[width=0.9\textwidth]{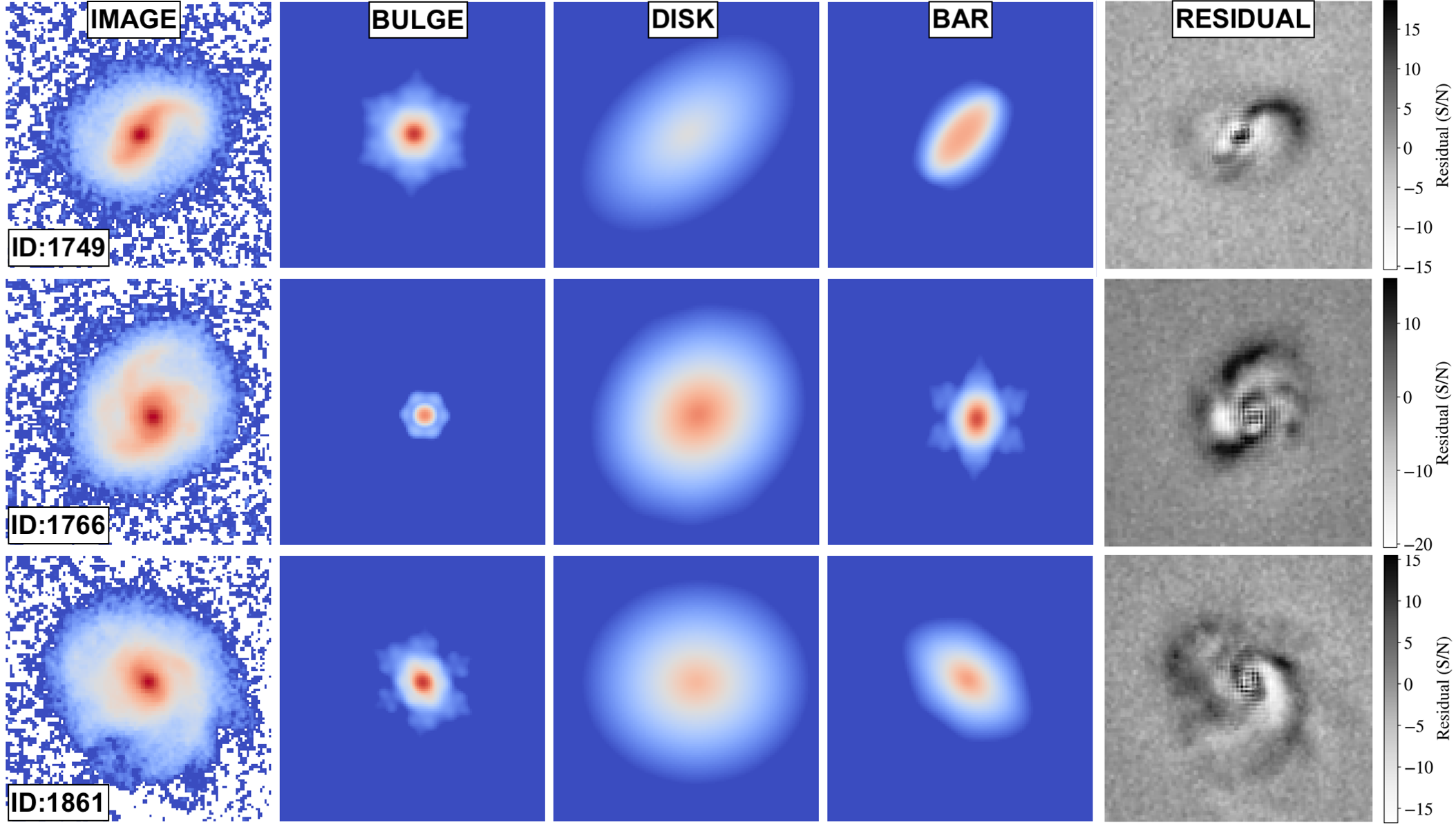}
\caption{2D morphological fitting of the rest-frame near-IR F444W images for the remaining three of the nine galaxies in our sample.}
\label{fig:2D_fit_comp_2}
\end{figure*}
The final step of our analysis aims to fully decompose the surface brightness distribution of our galaxies into bulge, bar, and disk contributions (Fig.~\ref{fig:2D_fit_comp_1}, \ref{fig:2D_fit_comp_2}). As in Sec.~\ref{subsec:spiral_mask}, we use spiral arm masks to minimize contamination.  Hence, these are the dominant features appearing in the residuals. Following the procedure in BK25, we first perform a bulge–disk dual Sérsic fit with the Python-based package \textsc{GALIGHT}\footnote{https://github.com/dartoon/galight}
 \citep{ding22}, which implements the forward-modeling tool \textsc{LENSTRONOMY}\footnote{https://github.com/lenstronomy/lenstronomy}
 \citep{birrer18, birrer21}. This approach provides access to the full posterior distribution of each fitted parameter, with optimization carried out using the Particle Swarm Optimizer \citep[PSO]{kennedy95}.  We use the F444W PSF generated using the software \textsc{PSFEx} \citep{bertin11} on the full COSMOS-Web mosaic from which the galaxy images have been extracted.  Furthermore, the error images are taken from the public COSMOS-Web database. These include the contributions of the corresponding weight images and the count-based Poisson noise.
 
The shapes of the bulge and disk from this initial fit are then used to initialize the corresponding components in the final bulge–bar–disk decomposition. The components of this final composite model are:

\begin{itemize}[leftmargin=*, noitemsep, topsep=1pt]
\item Bulge: Fixed $n=4$ Sérsic model, initialized with the flux (amplitude), size (effective radius), and shape ($e_1$, $e_2$) from the initial fit. 
\item Disk: Fixed $n=1$ Sérsic model, initialized from the initial disk fit. In some cases, however, the disk component in the initial fit included both the disk and bar (based on visual inspection). To prevent this, we constrain $e_1$ and $e_2$ to be within $10\%$ of the ellipse at $r_\mathrm{90}$ of the best-fit disk model. This provides a robust automated safeguard against bar contamination.
\item Bar: Variable $n=0.1-1.5$ Sérsic model, covering the full range of bar profiles observed in the S$^{4}$G sample at $z\sim0$ \citep{kim15}. We assume this range also applies at $z\sim1.5$. The bar phase is initialized from the ellipse-fitting results, while the initial $R_e$ is set to $2\times$ the bulge $R_e$.
\end{itemize}

It should be noted that the PSO method only allows for a model optimization, and does not provide uncertainties.  Hence, once the model converges, we run an additional MCMC sampling with the bulge and bar variables free. The disk flux is also left free, but its shape is fixed. Without this constraint, the disk can spuriously fit the bar, an issue we aim to avoid as discussed above. This step provides the error bars for the measured values used in later analysis.  

We check for dependence of our results on our choice  of the bulge $n$-value.  Hence, we repeat our analysis with bulge $n$ fixed to $2, 2.5,3,3.5$ since a high fraction of local bulges can have $n<4$ \citep{gao20}. Using $n=1$ results in degeneracy with the disk component and requires additional constraints which we have tried to avoid. We find that the bar profile (discussed throughout this paper) does not get influenced by our choice of bulge index within the range $n = 2-4$. The measured bar parameters remain within the corresponding uncertainties.

For this multi-component fitting, we have opted for the PSO-based \textsc{GALIGHT} rather than the more commonly used $\chi^2$-optimization-based \textsc{GALFIT} for a couple of reasons. Empirically, we find that the latter software is unable to converge for at least half of the sample. Meanwhile, the PSO-based approach results in successful convergence throughout our sample, and the resulting bar model shapes correspond well to the ellipse-fitting results in terms of position angle as well as visual shape. We conclude that the reason the PSO-based approach outperforms $\chi^2$ optimization is that the former explores the full range of the posterior distribution, whereas the latter follows local gradients in $\chi^2$, which also results in a heavy reliance on the initial guesses. However, we are unable to implement the multi-wavelength features of \textsc{GALFIT} that could have improved its performance, since we do not clearly detect the bars in shorter-wavelength images.

\subsection{The bar length uncertainty and its implication}
We fit the bars in our sample galaxies with a Sérsic profile. However, this does not allow us to measure bar length. Physically, the bar length is the maximum extent of the $x_\mathrm{1}$ orbits of the trapped stars making up the bar \citep{binney08}. In the local Universe, this typically corresponds to the major axis derived from ellipse fits to surface brightness isophotes (Sec.~\ref{subsec:ellipse_fit}), beyond which the bar surface brightness drops rapidly. For this reason, 2D Sérsic fitting of local bars often includes an additional parameter, $r_\mathrm{bar}$, the major-axis radius beyond which the profile is truncated to zero \citep{Kim14}. At $z\sim1.5$, however, we cannot include $r_\mathrm{bar}$ as a variable due to the combined effects of low signal-to-noise and flux confusion with other non-axisymmetric disk structures (discussed in BK25). For the same reason, we do not adopt the modified Ferrers profile \citep{laurikainen05}, which has a variable $\alpha$ describing the sharpness of the truncation. We plan to revisit this fitting procedure in future work with improved treatment of disk substructures.

Instead, we use bar lengths ($R_\mathrm{bar}$) from the ellipse-fit method: the SMA corresponding to the peak ellipticity. This follows the approach of \cite{kim21} for galaxies at $z \lesssim 0.84$. \cite{liang24} showed that this measure can overestimate bar length when the intrinsic length is $<$ PSF FWHM. However, detectability also drops sharply below this threshold. Since all bars in our sample have SMA values $\gtrsim$ PSF FWHM, we assume they are only detected because their intrinsic lengths are also $\gtrsim$ PSF FWHM. We therefore adopt an uncertainty of $10\%$ \citep[as measured by][for bars $\gtrsim$ PSF FWHM]{liang24}. We note, however, that the uncertainties in \cite{liang24} were derived from local galaxies after applying redshift-dependent systematic effects and disk scaling relations; intrinsic (sub-)structural properties at high redshift \citep{guo12, guo15, sattari23, kalita24, kalita25b, mercier25} were not considered. Investigating this effect is beyond the scope of this work.  Finally, we correct for the projection effects, by using the axis ratio of the disk ($q$) and the position angle difference between the major axes of the disk and the bar ($\theta$):
\begin{equation}
R_\mathrm{bar,corrected} = \left[(\frac{1}{q}\, R_\mathrm{bar}\, sin \,\theta)^2 + (R_\mathrm{bar}\,cos \,\theta)^2\right]^{1/2}
\end{equation}

The lack of truncation in our model may still bias some measurements, especially the bar Sérsic index. If truncation is present in the data but not modeled, the Sérsic index may be artificially lowered to account for the flux drop. To test this, we performed a second round of 2D fitting with a fixed-radius truncation, using the ellipse-derived bar length. The ellipse corresponding to the bar (Sec.\ref{subsec:ellipse_fit}) was used as the bar edge, beyond which the bar profile was masked. We then re-ran the MCMC sampling, starting from the previous best-fit values. All bar profile measurements, particularly the Sérsic index, remained consistent with the original values within their uncertainties. We therefore conclude that the flatness of the bar profiles indicated by the Sérsic index is robust, and consistent with the flat profiles observed both in the visual 1D major-axis profiles and the 1D fits (Sec.\ref{subsec:1d_fit}).

\section{Results}
The final compilation of the bar properties of the galaxies in our sample is provided in Table~\ref{tab:bar_properties}.  We now discuss these parameters in the following subsections.
\begin{table*}
\centering
\caption{Bar structural properties of the sample galaxies.}
\setlength{\tabcolsep}{6pt} 
\renewcommand{\arraystretch}{1.3} 
\begin{tabular}{*{10}{c}} 
\hline
\textbf{ID} & \textbf{RA} & \textbf{Dec} & \textbf{$z$} & \textbf{galaxy $\log(M_\mathrm{\star}/M_\mathrm{\odot})$} & \textbf{$R_\mathrm{\mathrm{bar}}$} & \textbf{$R_\mathrm{\mathrm{e,bar}}$} & \textbf{bar $n_\mathrm{\mathrm{1D}}$} & \textbf{bar $n_\mathrm{\mathrm{2D}}$} & \textbf{$Q_\mathrm{T}$} \\
 & [deg] & [deg] &  & [dex] & [kpc] & [kpc] &  &  &  \\
\hline
89    & 149.85575 & 2.13022 & 1.47840 & $11.17 \pm 0.04$ & 4.51 & 2.25 & $0.32 \pm 0.02$ & $0.17 \pm 0.03$ & $0.29 \pm 0.07$ \\
147   & 149.74583 & 2.12594 & 1.55634 & $11.06 \pm 0.05$ & 2.80 & 2.64 & $0.31 \pm 0.03$ & $0.59 \pm 0.16$ & $0.32 \pm 0.08$ \\
852   & 150.44154 & 2.12922 & 1.55744 & $11.38 \pm 0.06$ & 4.73 & 2.62 & --- & $0.17 \pm 0.11$ & $0.10 \pm 0.03$ \\
1147  & 149.85471 & 2.11511 & 1.48173 & $11.04 \pm 0.04$ & 4.96 & 1.98 & $0.57 \pm 0.03$ & $0.66 \pm 0.12$ & $0.26 \pm 0.07$ \\
1334  & 150.40292 & 2.40883 & 1.51410 & $10.87 \pm 0.10$ & 3.25 & 2.44 & $0.24 \pm 0.03$ & $0.20 \pm 0.07$ & $0.28 \pm 0.07$ \\
1704  & 149.77554 & 2.25164 & 1.67310 & $10.90 \pm 0.05$ & 2.42 & 2.17 & $0.62 \pm 0.02$ & $0.30 \pm 0.01$ & $0.07 \pm 0.02$ \\
1749  & 150.02500 & 2.35528 & 1.61099 & $10.82 \pm 0.06$ & 3.39 & 2.71 & $0.41 \pm 0.03$ & $0.39 \pm 0.20$ & $0.32 \pm 0.08$ \\
1766  & 149.87879 & 2.49983 & 1.67292 & $11.05 \pm 0.06$ & 2.44 & 0.89 & $0.69 \pm 0.02$ & $1.26 \pm 0.04$ & $0.25 \pm 0.06$ \\
1861  & 150.00108 & 2.32144 & 1.45947 & $10.86 \pm 0.07$ & 2.21 & 2.55 & $0.68 \pm 0.06$ & $1.36 \pm 0.01$ & $0.18 \pm 0.05$ \\
\hline
\end{tabular}
\label{tab:bar_properties}
\end{table*}

\subsection{Bar profile}
The first key property we study is the bar major-axis profile, which spans a range from flat to exponential. Using the 1D flux distribution along the bar ellipse major axis (Sec.~\ref{subsec:ellipse_fit}), we identify clear shoulders in 5/9 galaxies (Figs.~\ref{fig:profile_examples}, \ref{fig:profile_examples_2}), suggestive of already-present flat bars. This qualitative assessment is further supported by the 1D and 2D Sérsic fits. For comparison with the literature, we adopt the classification scheme of \cite{kim15}, which divides bars into flat ($n < 0.4$), intermediate ($0.4 < n < 0.8$), and exponential ($n > 0.8$). This highlights that bar surface brightness profiles are not strictly flat or exponential, but rather span a continuum.

\begin{figure}
\centering
\includegraphics[width=0.49\textwidth]{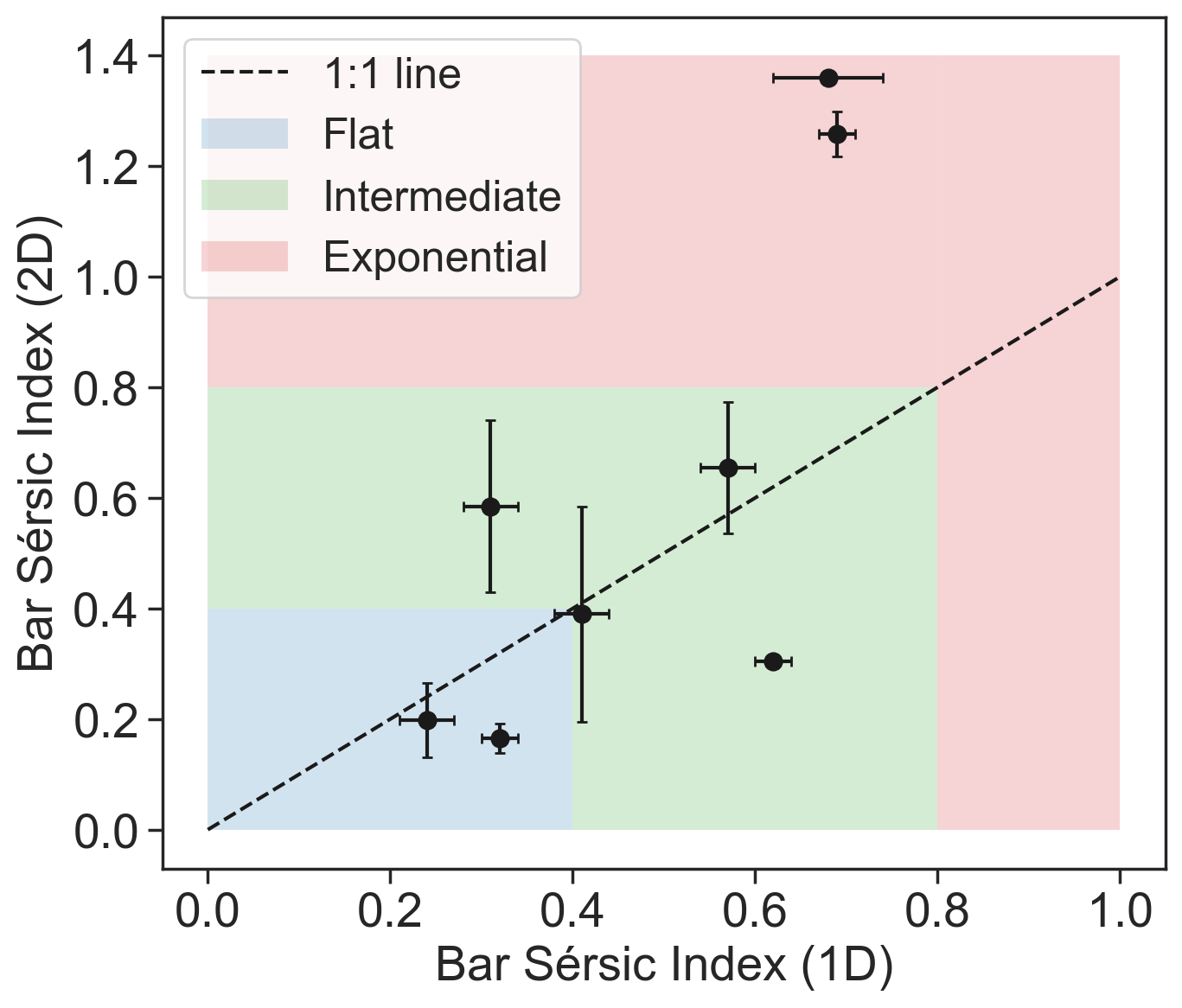}
\caption{Bar profiles in 1D and 2D: Sérsic index measurements for each galaxy in our sample. The x-axis shows the 1D fitting results (Sec.\ref{subsec:1d_fit}), while the y-axis shows the 2D fitting results (Sec.\ref{subsec:2d_fit}). The shaded regions indicate the classification scheme: flat ($n<0.4$), intermediate ($0.4<n<0.8$), and exponential ($n>0.8$).}
\label{fig:sersic_vals_bar}
\end{figure}

The 1D fits indicate that eight bars in our sample fall in the flat-to-intermediate regime ($n<0.8$), while the 1D fitting failed in the final one (ID: 852; albeit it clearly shows shoulders in the major-axis profile). However, the major-axis profile does not capture the full 2D flux distribution, and bar models may still be influenced by residual disk substructures, even after masking the primary spiral arms (Sec.~\ref{subsec:spiral_mask}). This is evident in some fits where the composite model converges to a nonphysical configuration, with the bar fitting large-scale outer-disk fluctuations. To mitigate this, we constrain the effective radius to $\lesssim 1/2 \times$ the bar ellipse major-axis length. After applying this constraint, the 1D Sérsic indices remain consistent with the visual major-axis profiles, with bars showing clear shoulders also exhibiting low Sérsic indices.

The 2D fits corroborate these results,  and agree well with the visual major-axis profiles as can be seen in Fig.~\ref{fig:profile_examples}, \ref{fig:profile_examples_2}. Sérsic indices from this analysis are generally consistent with the 1D values as can be seen in Fig.~\ref{fig:sersic_vals_bar}: for five of the eight bars with 1D Sérsic measurements, the indices agree within 0.1. These correspond to the bars with the lowest indices ($n<0.6$ in both 1D and 2D). We interpret this agreement as a consequence of flat-bar major-axis profiles being distinctly different from the exponential disk profiles. Among the remaining three bars, one has a 1D index of 0.65 and a 2D index of 0.35 (ID: 1704). The other two show 1D indices $\sim 0.7$ but 2D indices $>1.2$. We conclude that the closer a bar profile is to that of the disk, the more uncertain the 1D fits become.


\subsection{Bar lengths} \label{subsec:bar_length}

\begin{figure}
\centering
\includegraphics[width=0.49\textwidth]{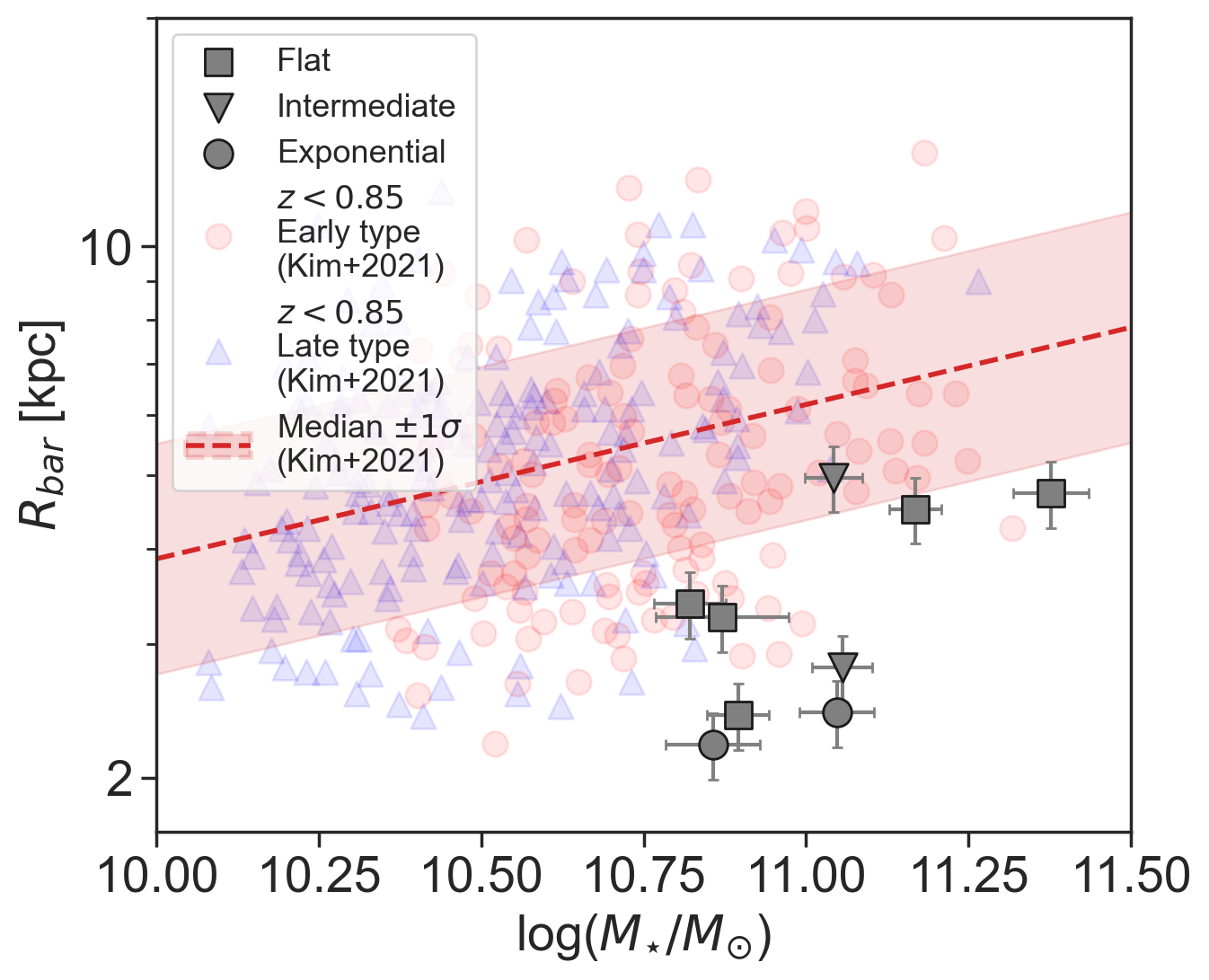}
\caption{$R_\mathrm{bar}$ vs. galaxy stellar mass. Also shown are values for late- and early-type barred galaxies at $0.2 < z < 0.85$ from \cite{kim21}, along with their median and $1\,\sigma$ scatter.}
\label{fig:R_bar_stellar_mass}
\end{figure}

\begin{figure}
\centering
\includegraphics[width=0.49\textwidth]{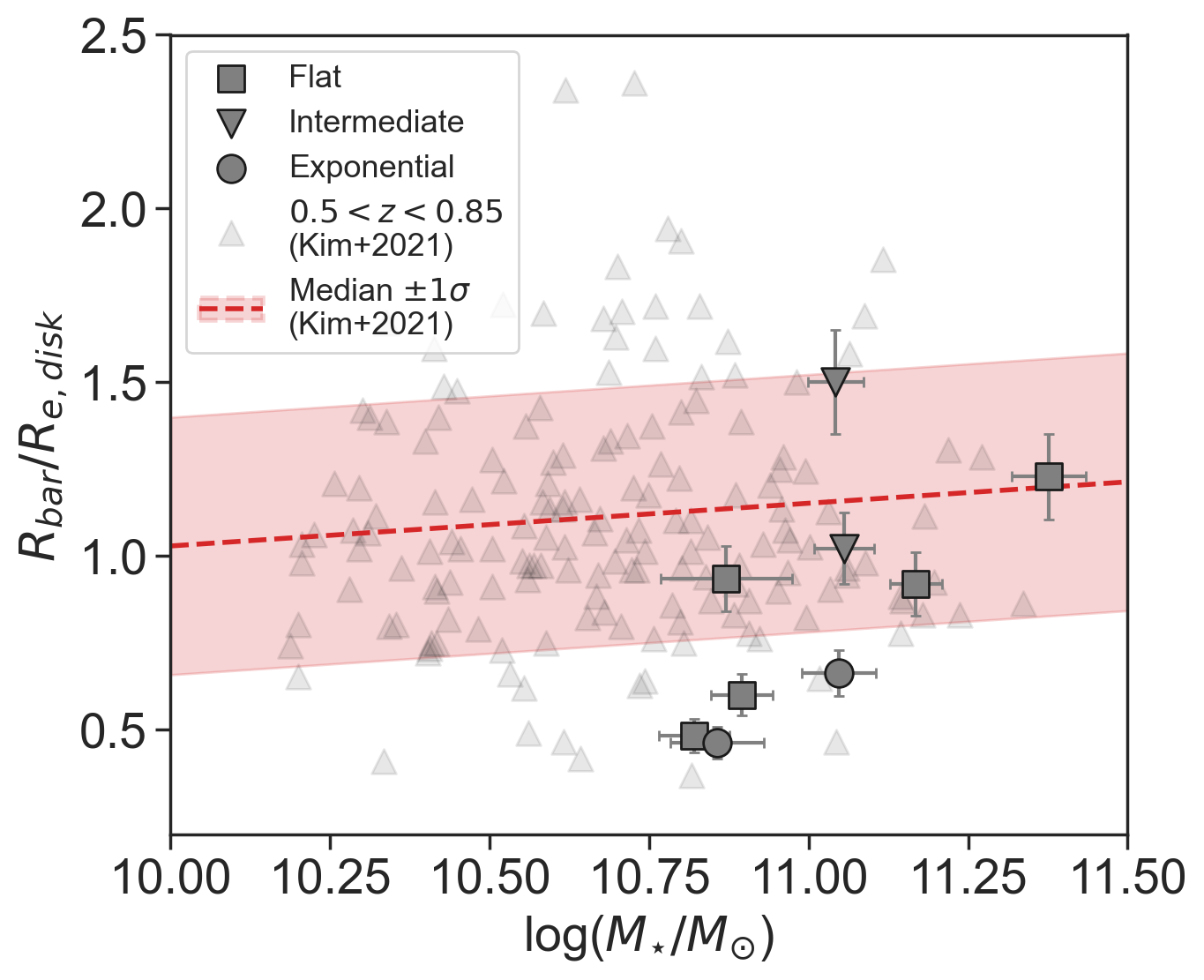}
\caption{$R_\mathrm{bar}$ normalized by the host disk effective radius vs. galaxy stellar mass for our sample. Gray points show galaxies at $0.5 < z < 0.85$ from \cite{kim21}. Their median and $1\,\sigma$ scatter are shown in red. }
\label{fig:R_bar_ratio_stellar_mass}
\end{figure}

\begin{figure}
\centering
\includegraphics[width=0.49\textwidth]{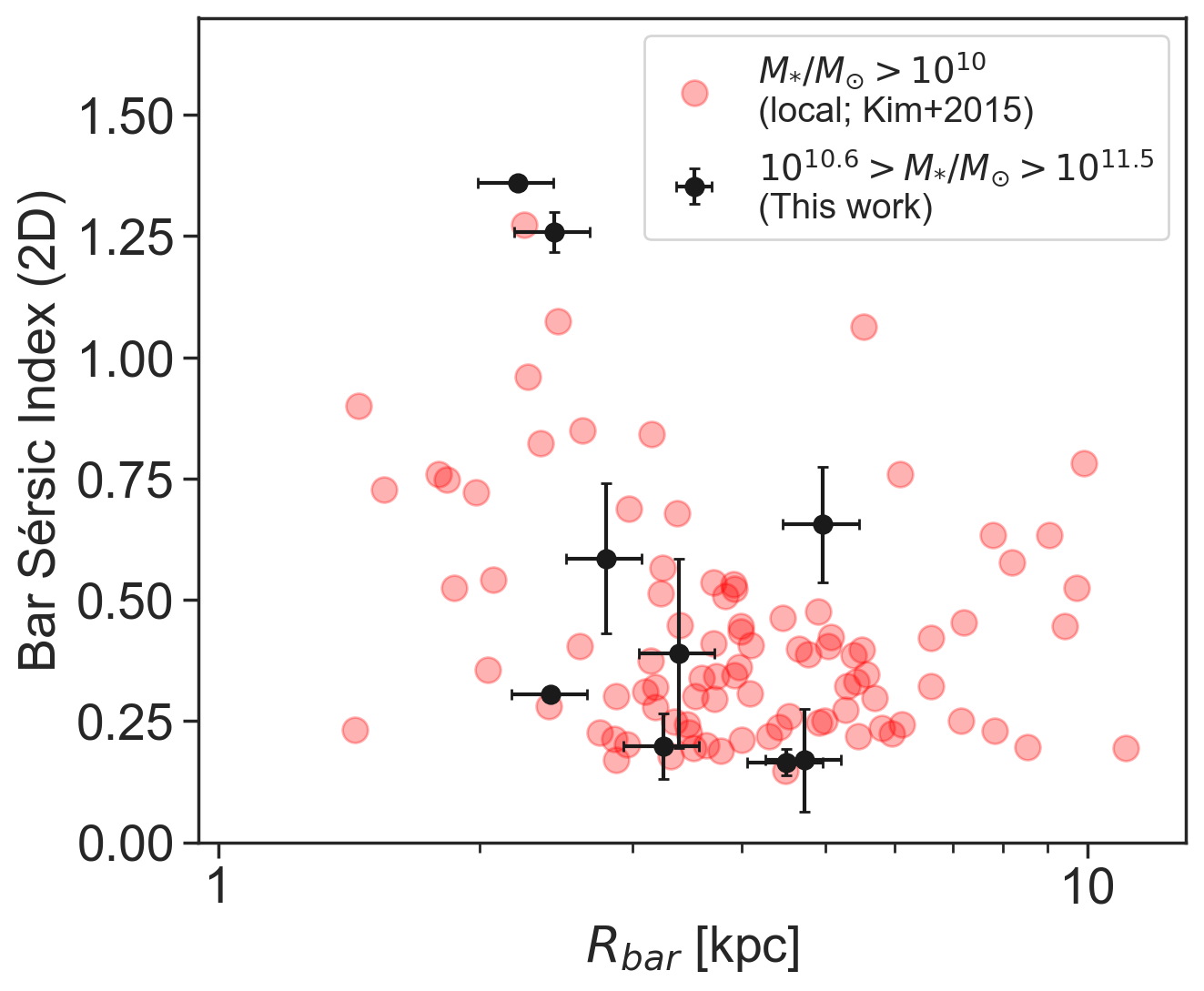}
\caption{Bar Sérsic index vs. $R_\mathrm{bar}$ for our sample. Values for massive ($M_\mathrm{*}>10^{10}\,\rm M_\mathrm{\odot}$) local galaxies from \cite{kim15} are also shown for comparison.}
\label{fig:Re_bar_sersic}
\end{figure}

\begin{figure}
\centering
\includegraphics[width=0.49\textwidth]{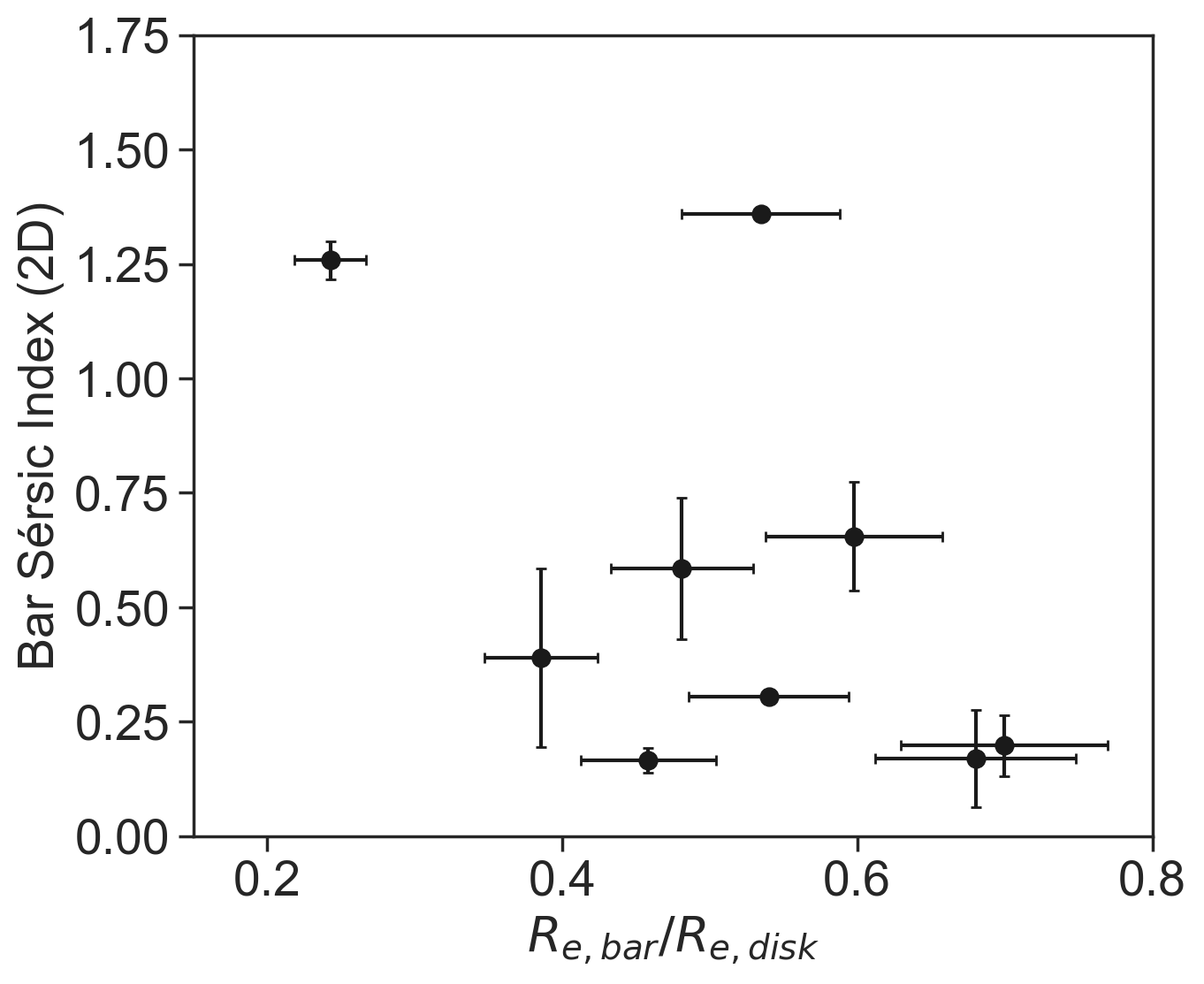}
\caption{Ratio of the effective radius of the bar to that of the host disk (x-axis) vs. bar Sérsic index (y-axis).}
\label{fig:Re_bar_Re_disk_sersic}
\end{figure}

The bar lengths ($R_\mathrm{bar}$) measured from ellipse fitting allow us to compare with bars at $z < 1$. As shown in Fig.\ref{fig:R_bar_stellar_mass}, our bars lie well below the median of the distribution observed at $0.2<z<0.85$ by \cite{kim21},  at $>1\,\sigma$ of their scatter.  A lower offset is seen in the normalized bar lengths (with the disk $R_e$ being the normalizing factor; Fig.\ref{fig:R_bar_ratio_stellar_mass}). Half of our values are within $\pm 1\,\sigma$ scatter around the median of their distribution, while the rest are below it. In both cases, we cannot determine any correlation with stellar mass from our data \citep[as expected in the local Universe;][]{erwin18, kim21}, due to the small sample size.

In Fig.~\ref{fig:Re_bar_sersic}, the two exponential bars are shorter by $\sim 1-2\,\rm kpc$ compared to the mean of the flatter bars in our sample. This trend between bar length and Sérsic index agrees with the distribution of $>10^{10}\,M_\mathrm{*}/M_\mathrm{\odot}$ galaxies at $0.2 < z < 0.85$ in \cite{kim21}.  This likely suggests that mature bars are longer, since the flattening of bars due to trapping stellar orbits also results in an increase in its length. However, if we plot the ratio of the effective radii ($R_\mathrm{e}$) of bar to disk(Fig.~\ref{fig:Re_bar_Re_disk_sersic}),  we find no clear correlation.  


\subsection{Bar flux ratio and axis ratio} \label{subsec:sbar_meas}
Theoretical studies conclude that bar growth is driven by the loss of angular momentum, which traps disk particles (stars) in the bar potential \citep{debattista00, athanassoula02, athanassoular03, valenzuela03, collier19a, collier19b, ansar24}.  In a more detailed sense, as the elongated bar (stellar) structure rotates with a constant angular velocity within a differentially rotating disk,  additional stars in the disk feel its gravitational effect.  Below the co-rotation radius (the radius at which the linear velocity of the bar and the disk match),  the stars are moving faster than the bar.  Some of these would lose angular momentum to the bar and align themselves with the elongated bar orbits. This reinforces the bar, thereby strengthening it. It should be noted however, that this process occurs primarily within the co-rotation radius and hence the bar length is lower than it \citep{aguerri98}. Gradually the bar slows down due to a net loss of angular momentum, which is expelled to the outer disk and the surrounding halo. As this happens, the co-rotation radius also increases, as the linear velocities of the bar and the disk will now match at a higher radius.  Hence, the bar is allowed to get longer, with greater ellipticity \citep{athanassoula02b}, as more stars align themselves with the bar.

\begin{figure}
\centering
\includegraphics[width=0.49\textwidth]{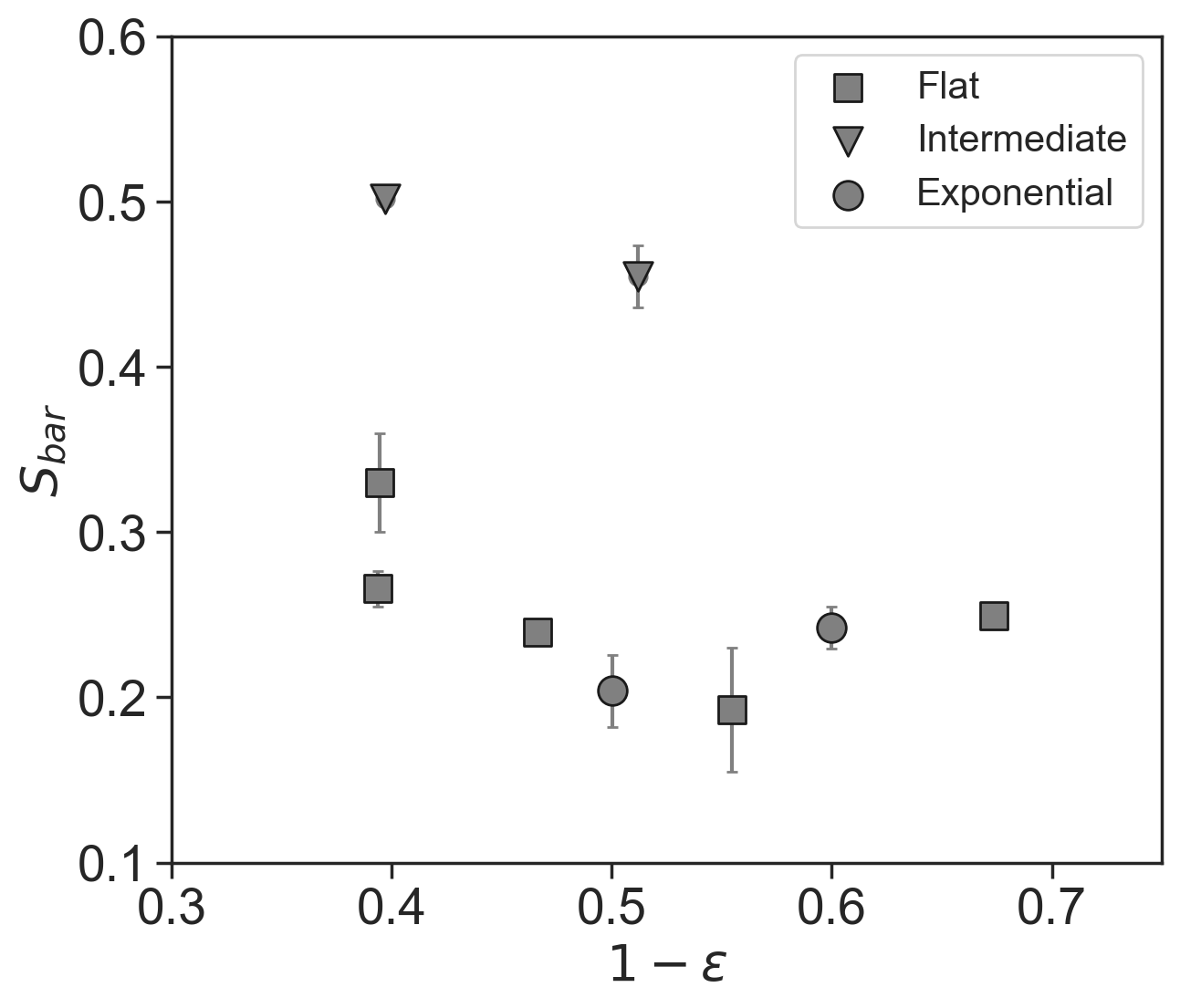}
\caption{The distribution of $S_\mathrm{bar}$ (defined in Eq.\ref{eq:sbar_definition}) and the axis ratio ($\equiv 1 - \epsilon$), both measured from the 2D fitting in Sec.~\ref{subsec:2d_fit}. We find a correlation coefficient $r = -0.42 \pm 0.18$.}
\label{fig:sbar_ellipticity}
\end{figure}

To further investigate the properties of the bars in our $z \sim 1.5$ sample, we examine the relation between bar flux ratio ($S_\mathrm{bar}$) and ellipticity.  The aim is to understand whether the bars we detect are increasing their ellipticity as they grow in surface brightness (due to addition of trapped stars). We define $S_\mathrm{bar}$ as:
\begin{equation} \label{eq:sbar_definition}
S_\mathrm{bar} = \frac{F_\mathrm{bar}}{F_\mathrm{total}}
\end{equation}
where the fluxes ($F$) are the model fluxes within the $3.6^{\prime\prime} \times 3.6^{\prime\prime}$ F444W images for each galaxy. Using the $e_1$ and $e_2$ values, we compute the axis ratio ($q \equiv 1 - \epsilon$). Uncertainties in both quantities are derived from the $1\sigma$ range of the posterior distributions in the MCMC sampling. We find a moderate correlation, $r = -0.42 \pm 0.18$ (Fig.~\ref{fig:sbar_ellipticity}), driven primarily by the two intermediate bars. The flat and exponential bars, by contrast, show similar $S_\mathrm{bar}$ values.

We choose this model-based parametrization rather than the classical $A_\mathrm{2}$ \citep[the Fourier amplitude of the $m=2$ component relative to $m=0$;][]{ohta90, laurikainen02, diazgarcia16, lee19} for three reasons. First, we avoid biases from imperfect deprojection: if the disk is not fully face-on, the $m=2$ amplitude is artificially boosted. Second, bulges are not always spherical, so even a correct deprojection can induce additional non-axisymmetries. This problem is amplified because the central regions of these $z \sim 1.5$ galaxies are only marginally resolved. Finally, Cosmic Noon galaxies contain substantial substructures such as clumps and spirals \citep{kalita25a, kalita24c}, which also contribute to the $m=2$ signal.

\subsection{Quantifying the effect of bars} \label{sec:q_bar_est}
Measurement of the relative flux contained within the bar (e.g., $S_\mathrm{bar}$) provides only a photometric estimate of bar strength, and does not directly quantify the physical \emph{effect} of the bar on its host galaxy. However, \cite{seidel15} found that spectroscopy-based kinematic torques induced by bars are well correlated with the maximum transverse-to-radial force ratio \citep[$Q_\mathrm{b}$;][]{combes81, buta01, laurikainen02, buta04, diazgarcia16, lee20}, which can be calculated directly from galaxy images. To date, $Q_\mathrm{b}$ has only been measured at $z < 0.8$ \citep{kim21}. At higher redshifts, low signal-to-noise ratios and the increasingly complex morphologies of galaxies, make such measurements challenging.

\begin{figure*}
\centering
\includegraphics[width=0.99\textwidth]{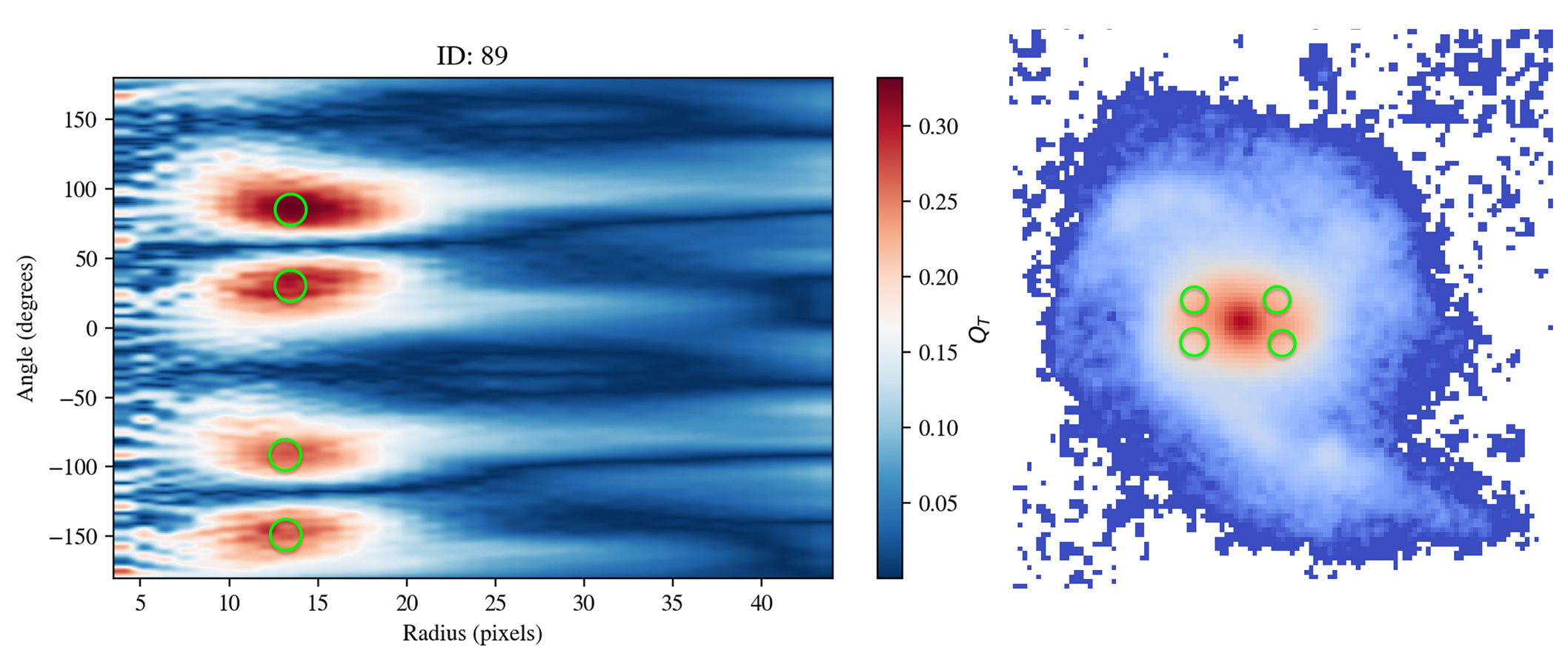}
\caption{Example of $Q_\mathrm{b}$ measurement. Left: distribution of $Q_T$ (Eq.~\ref{eq:qt_def}) in ($r,\phi$)-space for galaxy ID: 89. The four maxima (lime green circles) correspond to the ``corners" of the bar. Right: locations of these maxima overplotted on the galaxy image. The absolute peak of $Q_T$, max($Q_T$), is taken as $Q_\mathrm{b}$.}
\label{fig:qbar_example}
\end{figure*}

At $z>1$, any $Q_\mathrm{b}$ estimate would reflect the net effect of all non-axisymmetric substructures, since disks at these epochs host abundant clumps and spirals \citep{elmegreen05, elmegreen07, forster11, guo18, kalita24, kalita25b, kalita25a, kalita25c, mercier25,salcedo25}. Isolating the bar’s contribution is therefore challenging. To address this, we adopt a modified $Q_\mathrm{b}$ measurement that uses our 2D bulge–disk–bar models (Sec.~\ref{subsec:2d_fit}) instead of the direct galaxy images. This removes the contribution of other non-axisymmetric features. For the original procedure, we refer the readers to \cite{lee20}.

We first deproject the model image using the disk $e_\mathrm{1}$ and $e_\mathrm{2}$ values (which jointly give the axis ratio and the position angle), yielding a face-on surface brightness profile. This is converted to a mass surface density, $\Sigma(x,y)$, assuming a constant mass-to-light ratio derived from the total F444W flux and stellar mass \citep{kalita25a}.  This approximation may still result in radial variations since the central regions maybe highly obscured.  However, our choice of the longest band limits this effect. Moreover, we opt to not use a stellar mass map derived from a pixel-by-pixel SED fitting, since the JWST data only has 4 wavelength bands. We therefore find artificial fluctuations of the order of $\sim 10\%$ of the average value in the mass map, introduced due to fitting uncertainties. Given that our analysis depends on the local gradients, it is highly susceptible to such fluctuations. 

We assume the disk to be made up of multiple stacked thin disks,  with the integrated mass surface density $\Sigma(x,y)$. However, to account for the vertical profile, we introduce a thickness factor given as:
\begin{equation} \label{eq:thickness_factor}
\rho(z) = \frac{1}{2h_\mathrm{z}}\,e^{-|z/h_z|}
\end{equation}
where $h_z$ is the scale height.  Since our galaxies are nearly face-on (axis ratio $> 0.6$; BK25), we adopt $h_z = 0.8 \pm 0.3\,\rm kpc$, which is twice the scale height determined by \cite{lian24} using a JWST-based statistical study of edge-on galaxies. The factor of two is added since the $sech^2$ function used in their study drops twice as fast as the exponential function used in this work.  The thickness factor, when integrated over the full vertical extent, reduces to unity. This is essential as we are using the integrated $\Sigma(x,y)$ for the whole disk, rather than a surface density for each thin disk component we assume the disk is made up of.  Therefore the final density of the disk is given as:
\begin{equation}
\rho(x,y,z) = \Sigma(x,y)\,\rho(z)
\end{equation}
We then compute the gravitational potential by solving Poisson’s equation using Fast Fourier Transform:
\begin{equation}\label{eq:poissons}
\nabla^2 \Phi (x,y) = 4\pi\,G\,\rho(x,y,z).
\end{equation}
The resulting potential is converted from Cartesian to polar coordinates $(r,\phi)$ by re-gridding and linear interpolating. We calculate two potentials: $\Phi_\mathrm{0}(r)$ from the axisymmetric bulge+disk model, and $\Phi(r,\phi)$ which also includes the bar. Their derivatives give the mean radial force,
\begin{equation}
\left<F_R (r)\right> \equiv \frac{d \Phi_\mathrm{0}(r)}{dr},
\end{equation}
and the transverse force,
\begin{equation}
F_T (r,\phi) \equiv \frac{1}{r}\frac{\partial\Phi(r,\phi)}{\partial\phi}.
\end{equation}
The transverse-to-radial force ratio is then defined as
\begin{equation} \label{eq:qt_def}
Q_\mathrm{T}(r,\phi) = \frac{F_T(r,\phi)}{\left<F_R(r)\right>}.
\end{equation}

We map $Q_\mathrm{T}(r,\phi)$ over ($r,\phi$)-space (Fig.~\ref{fig:qbar_example}); its absolute peak defines $Q_\mathrm{b}$. For our sample, $Q_\mathrm{b}$ values lie between $\lesssim 0.10$ and $0.50$. The lower limit is uncertain, as polar gridding introduces local peaks $\lesssim 0.10$. Two galaxies (IDs: 852 and 1704) in our sample fall into this regime.

\section{Discussion}
As we have been able to characterize the morphological structure of bars in our sample at $z\sim1.5$, we will now discuss the implications.
\subsection{The presence of ``mature" bars} \label{subsec:mature_bars_disc}
Our work has leveraged the high-resolution imaging of JWST/NIRCam to determine the major-axis profiles of bars. Using both the 1D (Sec.\ref{subsec:1d_fit}) and 2D (Sec.\ref{subsec:2d_fit}) analyses, we find that seven of the nine barred galaxies have flat-to-intermediate ($n < 0.8$) profiles, while the other two lie in the exponential regime ($n > 0.8$). The implications of non-exponential profiles has been debated since the pioneering study of \cite{elmegreen85}. These profiles, generally referred to as flat bars \citep[and more recently subdivided into flat and intermediate by][]{kim15}, are mainly found in early-type systems in the local Universe \citep{elmegreen85, buamgart86, elmegreen96c, regan97}. This bias has been interpreted as evidence that flat bars evolve from exponential bars, growing longer and stronger in the process \citep{elmegreen89, kim15, lee19}. More recently, \cite{erwin23} showed that bar profile correlates more fundamentally with galaxy mass rather than Hubble type, while \cite{kruk18} suggested an evolutionary trend based on barred galaxy disks being redder, and hence lower in star-formation, than disks in unbarred galaxies. Direct evidence for profile evolution has also been found in N-body simulations: \cite{anderson22} showed that bars begin with exponential profiles and evolve into flat profiles in $\sim 1- 2\,$Gyr.
\begin{figure*}
\centering
\includegraphics[width=0.8\textwidth]{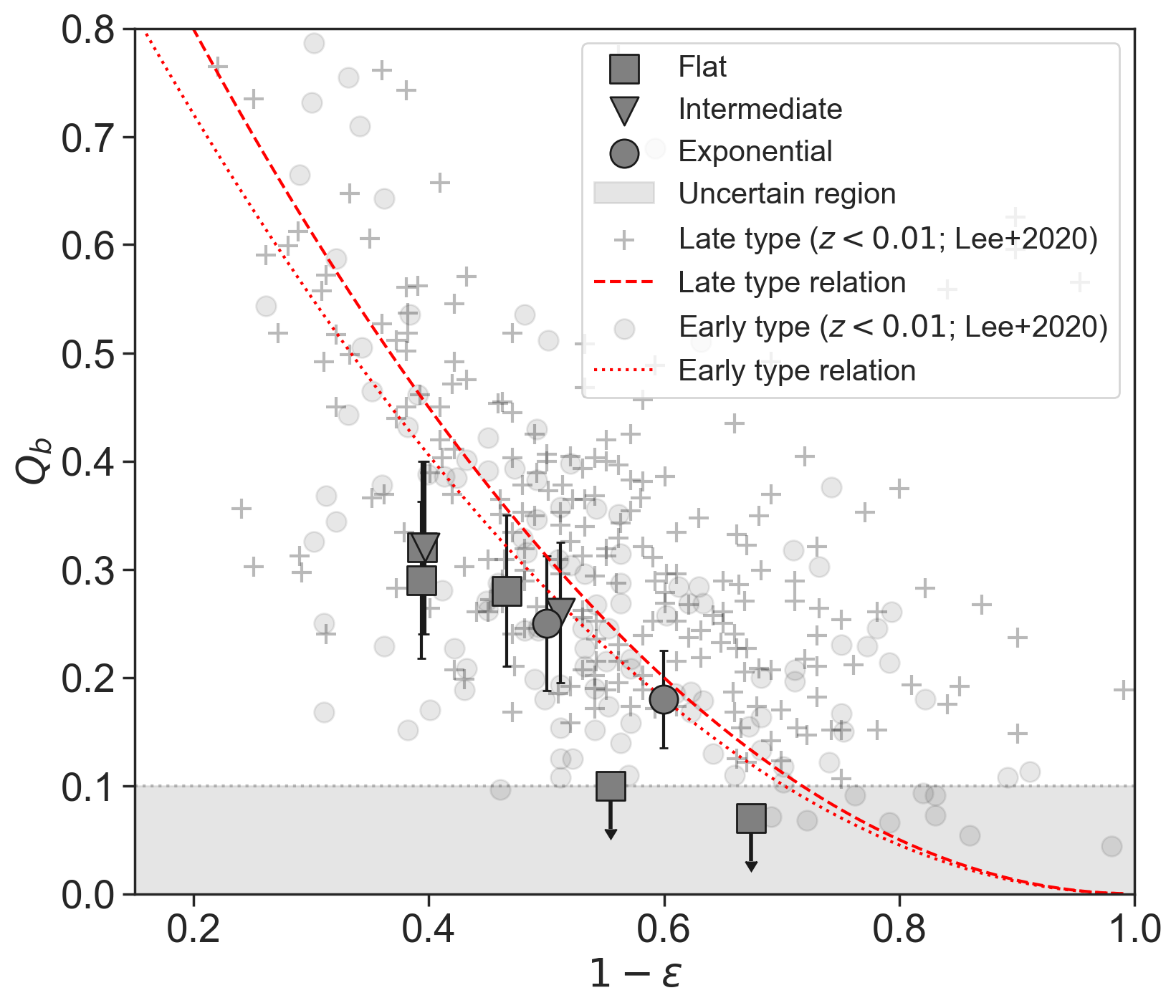}
\caption{Distribution of $Q_\mathrm{b}$ as a function of bar axis ratio, divided into three profile categories. Also shown is the \cite{lee20} distribution of local barred galaxies. The grey region denotes  $Q_\mathrm{b}$ values $\leq 0.1$,  the uncertainty threshold signifying the maximum systematic fluctuations of the measurement procedure.}
\label{fig:qbar_plot}
\end{figure*}

\begin{figure}
\centering
\includegraphics[width=0.49\textwidth]{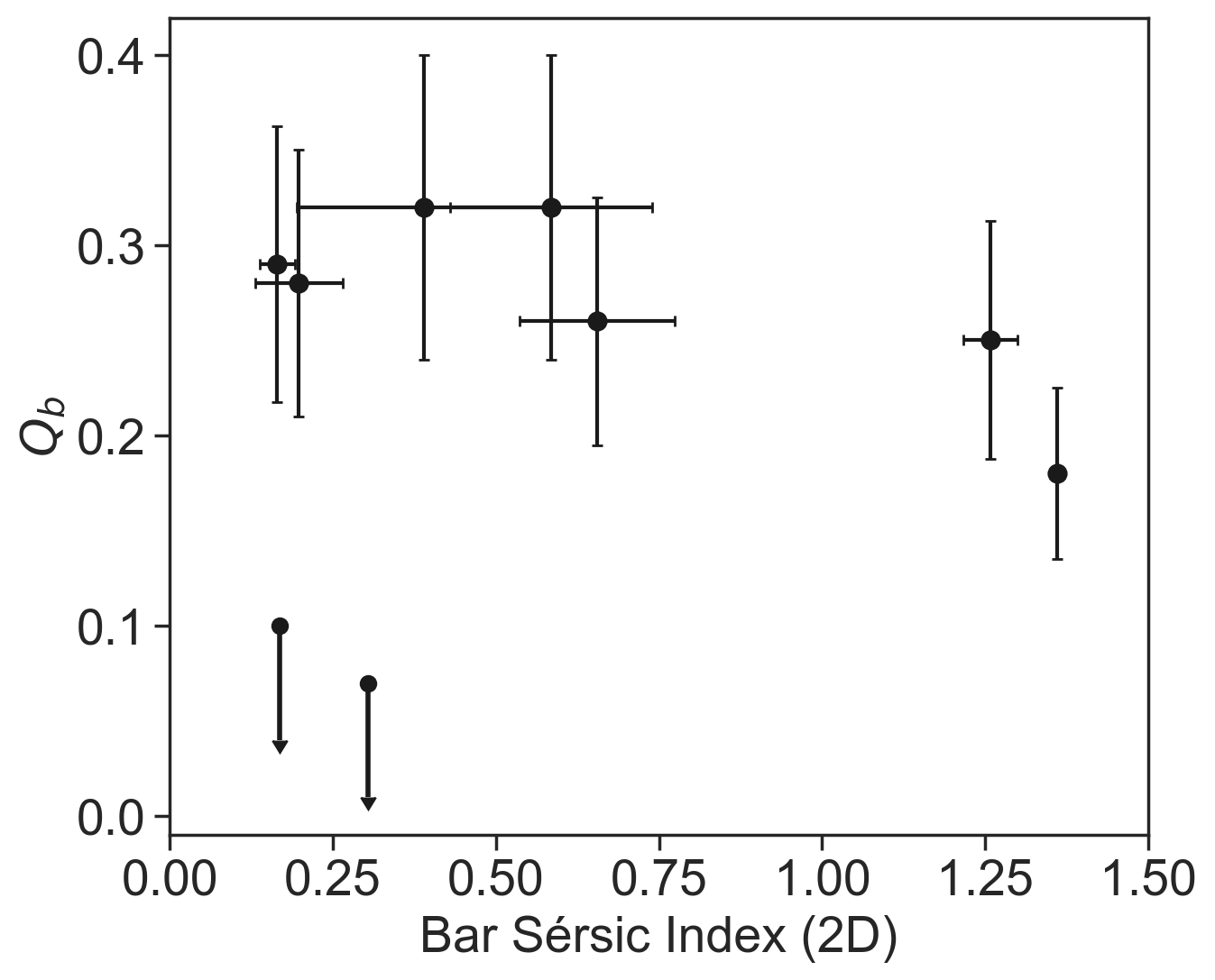}
\caption{$Q_\mathrm{b}$ vs. Sérsic index for the bars in our sample.}
\label{fig:qb_sersic}
\end{figure}

There are still gaps in our understanding of the formation and evolution of bars at $z>1$, where disks have higher dispersion in comparison to the local Universe \citep[by a factor $\sim 2$ based on ][]{ubler19}.  Simulations have found evidence that bars can still form in such conditions, albeit with weaker strengths \citep{ghosh23}.  These bars can also survive for $> 1-2\,\rm Gyr$ if the disk gas fractions are sufficiently low \citep[$<0.6$][]{blandhawthorn23}. There is however some uncertainty regarding the source of the initial $m=2$ purturbation. Although bars have been found to form spontaneously in the above mentioned N-body simulations, \cite{fragkoudi25} uses a suite of magneto-hydrodynamical cosmological zoom-in (Auriga) simulations to conclude that the intial $m=2$ purturbation is rather a result of mergers. This is also expected to result in bars that are already ``saturated'' in length and unable to grow further.  What this means for the bar profile is unclear however.  \cite{zheng25} similarly concluded that interaction-driven bars follow an accelerated evolutionary pathway. The only discussion regarding the effect of interaction on the bar profile has been provided in \cite{noguchi96}, where they suggest a rapid development of a flat profile using simple analytical models.

We do not observe any companions for eight of the galaxies in our sample, or any sign of clear tidal disruption (BK25). Furthermore, whether the bars in our sample are saturated at birth also cannot be verified with our bar length measurements (Sec.~\ref{subsec:bar_length}). The values predicted by the Auriga simulations for high-z saturated bars \citep[$>4\,\rm kpc$][]{fragkoudi25} are larger than our bar length estimates. However, our measurement do agree with the bar lengths from other JWST-based studies \citep[e.g.,][]{guo25} at $z \sim 1.5$. Such discrepancies between observations and simulations may reflect differences introduced by measurement methods rather than true physical disagreement, as the maximum-ellipticity we have adopted is known to underestimate the bar length \citep{wozniak95, athanassoula02b, laurikainen02, erwin03}.

The saturated bars are also expected to be longer than their local counterparts. Hence, more prudent is a comparison of our results to the barred galaxies at $z \sim 0.3-0.85$ in the S$_\mathrm{4}$G sample \citep{kim21}, which uses the same method of measurement. Our galaxies fall on the lower end of their distribution (Fig.~\ref{fig:R_bar_stellar_mass} for absolute lengths; Fig.~\ref{fig:R_bar_ratio_stellar_mass} for normalized lengths). While our study is limited by small-number statistics, other JWST-based surveys \citep{leconte24, guo25} contain similar numbers of galaxies per redshift bin, suggesting that our results are still representative of $z\sim 1.5$ galaxies. Hence, the shorter bars at this epoch in comparison to those at $z < 1$ does not agree with such bars being saturated at birth.  Therefore, we do not find any evidence that majority of the bars in our sample have been heavily influenced by interactions.  Therefore, the mature bars in our sample likely have developed as a result of secular evolution.

\subsection{The varying strength and impact of $z > 1$ bars} \label{subsec:bar_varying_strength}

Our first estimate of bar strength is the relative rest-frame near-IR bar flux with respect to the disk ($S_\mathrm{bar}$; Sec.\ref{subsec:sbar_meas}). We find that the maturity of a bar does not necessarily translate into higher relative surface brightness: flatter (and possibly longer) bars are not always brighter. A flux-based measure, however, does not capture the gravitational torque due to the bar potential, as it ignores bar morphology. Since longer and flatter bars exert larger torques \citep{lee20}, we instead use $Q_\mathrm{b}$ (Sec.\ref{sec:q_bar_est}) for a more complete quantification of the bar strength and its impact on the host disk.

Figure~\ref{fig:qbar_plot} shows that $Q_\mathrm{b}$ is inversely proportional to axis ratio, as expected for more elongated bars \citep{laurikainen02}. Our values follow the relation observed in the local Universe \citep{lee20}. While their study finds a small offset between early- and late-type galaxies, this difference is too subtle to make any comparisons to our limited sample. We also observe that our absolute $Q_\mathrm{b}$ values at $z \sim 1.5$ lie within the average and lower end of scatter of the $z < 0.01$ sample. We note that disks at these epochs are expected to be dynamically hotter and thicker. In our calculations, we adopt scale height range of $0.8 \pm 0.3\,\rm kpc$, and this range is propagated into the $Q_\mathrm{b}$ uncertainties. Thus the similarity in $Q_\mathrm{b}$ values indicate that despite different dynamical conditions, bars at high redshift exert comparable torques on their host disks as those in the local Universe. This result is consistent with \cite{kim21}, who found little redshift evolution in $Q_\mathrm{b}$ between $0.2 < z < 0.85$. Our values are also in agreement with theirs ($Q_\mathrm{b} = 0.3-0.25 \pm 0.15$).

Given the large intrinsic scatter in $Q_\mathrm{b}$ \citep{lee20, kim21}, the marginal decrease of $Q_\mathrm{b}$ with Sérsic index (Fig.~\ref{fig:qb_sersic}) for seven of the galaxies in our sample cannot be considered conclusive. Nonetheless, studies of local galaxies suggest that bars strengthen as they mature \citep{buamgart86, elmegreen89, kim15, lee19, kim21}, which is consistent with our findings. More puzzling are two remaining galaxies (IDs: 852 and 1704) that show flat bar profiles but with only upper limits estimates of $Q_\mathrm{b} \lesssim 0.10$, therefore falling below the uncertainty threshold (Sec.~\ref{sec:q_bar_est}). ID 1704 in particular has a distinct rectangular, plane-of-the-disk profile resembling the $4/1$ orbital family \citep{contopoulos88}, which typically emerges only in strong bars. We find that the low $Q_\mathrm{b}$ values in these galaxies arise from the combination of bar profile and flux. A dominant bulge could also lower $Q_\mathrm{b}$ by increasing $\left<F_\mathrm{R}\right>$ \citep{kim21}. But both galaxies have low bulge-to-total ratios (B/T $\sim 0.1$), at the lower end of our sample’s range ($0.05-0.4$). Thus, the question remains: \emph{how did these mature bars weaken?}

Another possibility is the addition of angular momentum from accreted gas \citep{bournaud05}. At $z>1$, galaxies experience high gas accretion rates, with the accretion onto the host halos scaling as $(1+z)^{2.25}$ \citep{genel08, goerdt10, dekel13}.  This could be playing a contributing role since ID: 852 has the second highest gas fraction\footnote{The gas mass is determined using metallicity dependent dust-to-gas ratio, where the dust mass is measured using multi-band SED fitting from the COSMOS2020 catalogue \citep{weaver22}. Further details will be provided in a follow-up paper} ($f_\mathrm{gas} = M_\mathrm{gas}/(M_\mathrm{gas}+M_\mathrm{\star}) = 0.64\pm0.12$) in our sample.  Finally, minor mergers can perturb stellar orbits and weaken bars \citep{martinezvalpuesta17, ghosh21}. Such interactions are expected to be more common at $z \gtrsim 1$ \citep{lotz11, mantha18}.  Once again, ID: 852 is the only galaxy in our sample that has a companion\footnote{we derive a photometric redshift of the companion using the software CIGALE and find it to be in agreement with the spectroscopic redshift of the galaxy. Furthermore, the location of the companion is aligned with one of the spiral arms. Hence, we claim that the structure is a physical companion.}  

\subsection{Bars tracing the evolution of high-z disks} \label{subsec:highz_disks}

The presence of flat mature bars in $z\sim 1.5$ indicates  that at least some disks had already undergone significant dynamical settling by this epoch,  albeit the exact time period required to establish the flatness is unclear. To estimate a timescale of evolution for the flat bars in our sample, placing the initial formation epoch at $z \sim 2.5-3.5$, we invoke the results of \cite{anderson22}. They find the shoulders, which give the characteristic flatness of a bar, appear after the bar evolves for $\sim1-2\,\rm Gyr$ in a secular disk. The models used are purely N-body, without the inclusion of gas and star-formation. The effects of tidal interactions have not been studied. They do, however, study disks with a range of thicknesses up to $\sim 900\,\rm pc$, hence covering the range observed at $z>1$ \citep{lian24}. With these caveats in mind, we still emphasize that the detection of such profiles provides a constraint on the conditions of early disks of our Universe at $z \gtrsim 2$.

The detection of bars up to $z\sim 3.5$ with JWST \citep[e.g.,][]{guo23,leconte24,guo25} has already challenged the classical expectations of bars being non-existent beyond $z>1$ \citep{sheth12}.  More recently, \cite{desafreitas25} have also found that bars in the local Universe might have existed well before $z\sim1$ based on stellar age estimates. These disks with high gas fractions \citep{magdis12, genzel15,saintonge16,tacconi18} and high turbulence-driven (ionized) gas velocity dispersion \citep[found to be $2-5\,\times$ higher than that in the local Universe]{labbe03,genzel11,wisnioski15,forster18} were expected not to be conducive for bar formation. More recently, however, \cite{ubler19} have presented an extensive study of 535 star-forming disk galaxies, where they use the KMOS instrument on the Very Large Telescope \citep{sharples13} to determine the ionized gas velocity dispersion to be $30-45,\rm km s^{-1}$ at $z\sim 1-2.5$. This is $\sim 2\times$ the values observed in the local Universe \citep[$24 \pm 5\,\rm km s^{-1}$ in the GHASP sample;][]{epinat10}.

Meanwhile, the molecular gas velocity dispersion is found to be $\sim 10-15\,\rm km s^{-1}$ lower, including in galaxies at $z>1$. Hence the molecular gas located in star-forming disks is dynamically decoupled from the low-mass ionized gas components. Meanwhile, the structure of bars and spirals is affected by the molecular and stellar components of the disk \citep[the stars inherit the velocity dispersion of the molecular gas;][]{vanDonkelaar22}, rather than the ionized gas \citep{jog84, romeo13}. Hence the velocity dispersion directly influencing the formation and evolution of secular structures like bars and spirals would rather be limited to $\sim 30\,\rm km s^{-1}$ at least up to $z\sim 2.5$. Hence, the extreme conditions that were thought to prevent the formation of bars may not be that extreme. 

Our results, however, do not simply indicate the presence of bars at $z>1.5$, but also indicate their long-term survival. This conclusion extends the discussion of bars at high-z beyond the emergence of the $m=2$ modes, and into sustained evolution for over $\sim 1-2\,\rm Gyr$. Using the \textsc{AGAMA/RAMSES} hydrodynamical N-body simulations, \cite{blandhawthorn24} explore the formation and evolution of bars in high-z turbulent disks. They find that galaxies with $f_\mathrm{gas} < 0.6$ can have bars forming and evolving for at least $2\,\rm Gyr$. It is hence worth noting that seven out of the nine galaxies in our sample have $f_\mathrm{gas} < 0.6$. Out of the two exceptions, ID: 852 has a weak bar, as has been discussed in the previous section. The other exception is ID: 1749, which has a strong bar. But one could speculate that the high gas fraction could be due to a more recent accretion of gas and has not influenced the bar for the majority of its lifetime. Nevertheless, larger statistics would be required to establish a firm connection between gas fractions and bar strengths.

Finally, \cite{fujii18} provides a theoretical framework on predicting the disk dominance after disk settling\footnote{a disk is considered settled when the stellar and gas components are in equilibrium with the dark matter halo and the stellar bulge}, at the epoch of bar formation. \cite{blandhawthorn23} verifies the relation using 3D hydro simulations. They conclude that the timescale of bar formation rises exponentially with the decrease in the disk dominance parameter, $f_\mathrm{disk}$,  which is defined as the square of the ratio of the circular velocities for the disk and the whole galaxy, measured at $2.2 \times$ the scale length of the disk \citep[Eq.~11 in ][]{fujii18}. Once formed, they remain stable and evolve for at least $\sim 1-2\, \rm Gyr$. Hence, the mature bars in our study verify this prediction, since we know these bars are not simply transient $m=2$ perturbations. Furthermore, with the expectation of these bars forming at $z \sim 2.5-3.5$, we can also deduce that the $f_\mathrm{disk}$ at that epoch would need to be $>0.4-0.5$ \citep[based on Fig.~7 in][]{blandhawthorn23}. It should be noted that direct measurement of $f_\mathrm{disk}$ requires expensive high resolution IFU spectroscopic observations \citep[e.g.,][]{price21}. Therefore, quantification of the bar profile allows us to reliably determine long-lived (mature) bars which can be used to indirectly estimate parameters like $f_\mathrm{disk}$, an estimation of the disk dominance in a galaxy.

\section{Summary}

In this work, we explore the nature of bars in a sample of nine massive galaxies ($M_\mathrm{*}/M_\mathrm{\odot} = 10^{10.6-11.4}$) at $z_\mathrm{spec} \sim 1.5$. We use rest-frame near-IR JWST/NIRCam F444W imaging from the COSMOS-Web survey \citep{casey23}. Leveraging the high spatial resolution, we perform a suite of morphological analyses to derive a range of bar properties:
\begin{itemize}[leftmargin=*, noitemsep, topsep=1pt]
\item Ellipse fitting to surface brightness isophotes, used to identify bars and measure their lengths.
\item Bulge–bar–disk model fitting to the major-axis surface brightness profile, which provides a visualization of the bar profile and initial conditions for the 2D analysis.
\item Bulge–bar–disk model fitting to the images, yielding bar Sérsic indices (as a proxy for the major-axis profile), effective radii, and bar-to-host flux ratios.
\item Estimation of the radial-to-transverse force ratio ($Q_\mathrm{b}$) from the best-fit 2D models, providing a measure of bar strength and the gravitational impact on the host galaxy.
\end{itemize}

\noindent
From these analyses, we obtain key results that reveal the nature of galactic bars at Cosmic Noon:
\begin{itemize}[leftmargin=*, noitemsep, topsep=1pt]
\item We find that five of the nine galaxies host flat, mature bars with $n<0.4$. Of the remaining four, only two bars have exponential profiles ($n>0.8$), indicating young bars. Hence, the bars at $z\sim 1.5$ have survived long enough ($\gtrsim 1-2\,\rm Gyr$, based on N-body simulations) to develop flat profiles (Sec.~\ref{subsec:mature_bars_disc}).
\item The mature bars are longer in absolute length than the two exponential bars, suggesting that bar lengthening occurs in parallel with profile flattening.
\item We measure the transverse-to-radial force ratio, $Q_\mathrm{b}$, for seven of the nine bars in the sample, providing an estimate of bar strength. We find the $Q_\mathrm{b}$ values at $z\sim 1.5$ to be comparable to those of local bars.
\item For the bars with measured $Q_\mathrm{b}$, we find a marginal decrease of the value with the bar Sérsic index (Sec.~\ref{subsec:bar_varying_strength}), consistent with the expectation that bars grow stronger as they mature.
\item However, the two galaxies with the lowest $Q_\mathrm{b}$ values (upper limits $\lesssim 0.1$) also have flat profiles. We conclude that these bars may have weakened due to gas accretion and/or tidal disruptions.
\item Through comparison with simulations, we conclude that the mature bars at $z\sim 1.5$ suggest the presence of settled, dominant disks by $z\sim 2.5-3.5$ (Sec.~\ref{subsec:highz_disks}).
\end{itemize}

In conclusion, we demonstrate that the detailed characterization of $z>1$ bar morphology, made possible by JWST, can reveal important insights into the nature and longevity of such bars, and by association, the host disks.

\begin{acknowledgments}
The authors are very grateful for the valuable comments by the anonymous referee. BSK was supported by the China Postdoctoral Science Foundation under Grant Number 2025M773186. LCH was supported by the National Science Foundation of China (12233001) and the China Manned Space Program (CMS-CSST-2025-A09).  This work was made possible by utilizing the CANDIDE cluster at the Institut d’Astrophysique de Paris, which was funded through grants from the PNCG, CNES, DIM-ACAV, and the Cosmic Dawn Center and maintained by Stephane Rouberol.  This work is based on observations made with the NASA/ESA/CSA James Webb Space Telescope. The data were obtained from the Mikulski Archive for Space Telescopes at the Space Telescope Science Institute, which is operated by the Association of Universities for Research in Astronomy, Inc., under NASA contract NAS 5-03127 for JWST. These observations are associated with program \#1727. The specific
observations analyzed in this work can be accessed via DOI: \href{https://doi.org/10.17909/p4e8-bz21}{10.17909/p4e8-bz21}.  BSK would like to thank Jinyi Shangguan for the valuable discussions. Finally, BSK thanks Xueling Wang for the constant support, without which this work would not have been possible.
\end{acknowledgments}

\begin{contribution}


BSK was responsible for most of the aspects ofJDS, ED,MD and AP provided valuable comments during the compilation of the results this manuscript.  LCH and FB provided key insights into the interpretation of the results.  .  XD and SY provided help with components of the code that was used for the analysis. Finally, all other authors provided valuable comments during the preparation of the manuscript.  


\end{contribution}

%
\facilities{JWST}

\software{astropy \citep{astropy22},  
          GALIGHT \citep{ding22}
          }



\bibliography{main}{}
\bibliographystyle{aasjournalv7}



\end{document}